\documentclass{emulateapj}  
\usepackage{courier,graphicx,epsfig,natbib,url,twoopt,color}
\usepackage{hyperref} 
\usepackage{pdfcomment}       
\usepackage{acronym}          
\hypersetup{
  colorlinks=true,  
  urlcolor=blue,  
  linkcolor=red  
}
\newdateformat{mydate}{\THEYEAR,\monthname[\THEMONTH], \THEDAy}

\makeatletter
\long\def\frontmatter@title@above{
  \vspace*{-\headsep}\vspace*{\headheight}
   Accepted by {\sc The Astrophysical Journal} 
  \par\vspace*{-\baselineskip}\vspace{12mm}
  }

\makeatother

\def\# #1\par{\par\mbox{}\\ \noindent{\color{red}\small $\sharp$ #1}\\} 

\bibpunct{(}{)}{;}{a}{}{,}    
\makeatletter
 \newcommandtwoopt{\citeads}[3][][]{%
   \nonstopmode
   \href{http://adsabs.harvard.edu/abs/#3}%
        {\def\hyper@linkstart##1##2{}%
         \let\hyper@linkend\@empty\citealp[#1][#2]{#3}}
   \biblink{#3}{\href{http://adsabs.harvard.edu/abs/#3}{ADS}}%
   \errorstopmode}            
 \newcommandtwoopt{\citepads}[3][][]{%
   \nonstopmode
   \href{http://adsabs.harvard.edu/abs/#3}%
        {\def\hyper@linkstart##1##2{}%
         \let\hyper@linkend\@empty\citep[#1][#2]{#3}}
   \biblink{#3}{\href{http://adsabs.harvard.edu/abs/#3}{ADS}}%
   \errorstopmode}            
 \newcommandtwoopt{\citetads}[3][][]{%
   \nonstopmode
   \href{http://adsabs.harvard.edu/abs/#3}%
        {\def\hyper@linkstart##1##2{}%
         \let\hyper@linkend\@empty\citet[#1][#2]{#3}}
   \biblink{#3}{\href{http://adsabs.harvard.edu/abs/#3}{ADS}}%
   \errorstopmode}            
 \newcommandtwoopt{\citeyearads}[3][][]{%
   \nonstopmode
   \href{http://adsabs.harvard.edu/abs/#3}%
        {\def\hyper@linkstart##1##2{}%
         \let\hyper@linkend\@empty\citeyear[#1][#2]{#3}}
   \biblink{#3}{\href{http://adsabs.harvard.edu/abs/#3}{ADS}}%
   \errorstopmode}            
\makeatother

\newcount\longrefs
\def\aap{\ifnum\longrefs=1 {Astron.\ Astrophys.}\else 
                           {A\hbox{\rm \&}A}\fi}
\def\aapr{\ifnum\longrefs=1 {Astron.\ Astrophys.\ Rev.}\else 
                            {A\hbox{\rm \&}AR}\fi}
\def\aaps{\ifnum\longrefs=1 {Astron.\ Astrophys.\ Suppl.}\else 
                            {A\hbox{\rm \&}A Suppl.}\fi}
\def\actaa{\ifnum\longrefs=1 {Acta Astronomica}\else
                            {Acta Astron.}\fi}
\def\aipcs{\ifnum\longrefs=1 {Am.\ Inst.\ Phys.\ Conf.\ Series}\else
                             {AIP Conf.\ Ser.}\fi}
\def\aj{\ifnum\longrefs=1 {Astron.\ J.}\else 
                          {AJ}\fi} 
\def\ao{\ifnum\longrefs=1 {Applied Optics}\else 
                           {Appl.\ Opt.}\fi} 
\def\aspcs{\ifnum\longrefs=1 {Astron.\ Soc.\ Pacific Conf.\ Series}\else 
                           {ASP Conf.\ Ser.}\fi} 
\def\apj{\ifnum\longrefs=1 {Astrophys.\ J.}\else 
                           {ApJ}\fi} 
\def\apjl{\ifnum\longrefs=1 {Astrophys.\ J. Lett.}\else 
                            {ApJL}\fi} 
\def\aplett{\ifnum\longrefs=1 {Astrophys.\ J. Lett.}\else 
                            {ApJ}\fi} 
\def\apjs{\ifnum\longrefs=1 {Astrophys.\ J. Suppl.}\else 
                            {ApJS}\fi}
\def\apss{\ifnum\longrefs=1 {Astrophys.\ and Space Science}\else 
                            {Astrophys.\ Space Sci.}\fi}
\def\araa{\ifnum\longrefs=1 {Ann.\ Rev.\ Astron.\ Astrophys.}\else 
                            {ARA\hbox{\rm \&}A}\fi}
\def\azh{\ifnum\longrefs=1 {Astronomicheskii Zhurnal}\else 
                            {Astron.\ Zhur.}\fi}
\def\baas{\ifnum\longrefs=1 {Bull.\ Am.\ Astron.\ Soc.}\else 
                            {BAAS}\fi}
\def\bain{\ifnum\longrefs=1 {Bull.\ Astronom.\ Institutes Netherlands}\else
                            {Bull.\ Astr.\ Inst.\ Neth.}\fi}
\def\cjaa{\ifnum\longrefs=1 {Chinese Jour.\ Astron.\ Astrophys.}\else 
                            {Chin.\ J.\ A\&A}\fi}
\def\gca{\ifnum\longrefs=1 {Geochim.\ Cosmochim.\ Acta}\else 
                           {Geochim.\ Cosmochim.\ Acta}\fi}
\def\grl{\ifnum\longrefs=1 {Geophys.\ Res.\ Lett.}\else 
                           {Geoph.\ Res.\ Lett.}\fi}
\def\iaucirc{\ifnum\longrefs=1 {IAU Circulars}\else 
                          {IAU Circ.}\fi}
\def\icarus{\ifnum\longrefs=1 {Icarus}\else 
                          {Icarus}\fi}
\def\ip{\ifnum\longrefs=1 {in press}\else 
                          {in press}\fi}
\def\jcap{\ifnum\longrefs=1 {Jour.\ Cosmology Astropart.\ Phys.}\else 
                          {JCAP}\fi}
\def\jgr{\ifnum\longrefs=1 {J.\ Geophys.\ Res.}\else 
                           {J.\ Geophys.\ Res.}\fi}  
\def\jrasc{\ifnum\longrefs=1 {J.\ Royal Astron.\ Soc.\ Canada}\else 
                           {JRAS Can.}\fi}  
\def\memsai{\ifnum\longrefs=1 {Mem.~Soc.~Astron.~Italiana}\else
                              {MmSAI}\fi}
\def\mnras{\ifnum\longrefs=1 {Mon.\ Not.\ Roy.\ Astron.\ Soc.}\else 
                             {MNRAS}\fi} 
\def\na{\ifnum\longrefs=1 {New Astronomy}\else 
                           {New Astron.}\fi}
\def\nar{\ifnum\longrefs=1 {New Astronomy rev.}\else 
                           {New Astron.\ Rev.}\fi}
\def\nat{\ifnum\longrefs=1 {Nature}\else 
                           {Nat}\fi}
\def\pasa{\ifnum\longrefs=1 {Pub.\ Astron.\ Soc.\ Australia}\else 
                            {PASA}\fi} 
\def\pasj{\ifnum\longrefs=1 {Pub.\ Astron.\ Soc.\ Japan}\else 
                            {PASJ}\fi} 
\def\pasp{\ifnum\longrefs=1 {Pub.\ Astron.\ Soc.\ Pacific}\else 
                            {PASP}\fi} 
\def\physscr{\ifnum\longrefs=1 {Physica Scripta}\else 
                            {Phys.\ Scrip.}\fi} 
\def\planss{\ifnum\longrefs=1 {Planetary \& Space Science}\else 
                            {Plan. \& Space Sci.}\fi} 
\def\procspie{\ifnum\longrefs=1 {Proc.\ SPIE}\else 
                            {Proc.\ SPIE}\fi} 
\def\qjras{\ifnum\longrefs=1 {Quarterly J.\ Royal Astron.\ Soc.}\else 
                            {QJRAS}\fi} 
\def\rmxaa{\ifnum\longrefs=1 {Revista Mexicana de Astron.\ y Astrofys.}\else 
                            {RMxAA}\fi} 
\def\sa{\ifnum\longrefs=1 {Soviet Astron..}\else 
                               {Sov.\ Astron.}\fi}
\def\skytel{\ifnum\longrefs=1 {Sky \& Telescope}\else 
                            {Sky \& Tel.}\fi} 
\def\solphys{\ifnum\longrefs=1 {Solar Phys.}\else 
                               {SoPh}\fi}
\def\sovast{\ifnum\longrefs=1 {Soviet Astronomy}\else 
                               {Sov.\ Ast.}\fi}
\def\ssr{\ifnum\longrefs=1 {Space Science Rev.}\else 
                               {Space\ Sci.\ Rev.}\fi}
\def\zap{\ifnum\longrefs=1 {Zeitschr.\ f.\ Astrophysik}\else
                               {Z.\ Astrophys.}\fi}

\makeatletter
\newcommand{\bibnote}[2]{\@namedef{#1note}{#2}}
\newcommand{\biblink}[2]{\@namedef{#1link}{#2}}
\makeatother

\def\acdef#1{\acl{#1} ({#1})}     
\newacro{AA}{Astronomy \& Astrophysics}
\newacro{ADS}{Astrophysics Data System}
\newacro{AIA}{Atmospheric Imaging Assembly}
\newacro{AO}{adaptive optics}
\newacro{ApJ}{Astrophysical Journal}
\newacro{AR}{active region}
\newacro{BFI}{Broad-band Filter Imager}
\newacro{CE}{coronal equilibrium}
\newacro{CfA}{Center for Astrophysics}
\newacro{CME}{coronal mass ejection}
\newacro{CRD}{complete redistribution}
\newacro{CRISP}{CRisp Imaging SpectroPolarimeter}
\newacro{CRISPEX}{CRisp SPectral EXplorer}
\newacro{CS}{coherent scattering}
\newacro{DEM}{Differential Emission Measure}
\newacro{DKIST}{Daniel K. Inouye Solar Telescope}
\newacro{DLR}{Deutsches Zentrum f\"ur Luft- und Raumfahrt}
\newacro{DOT}{Dutch Open Telescope}
\newacro{DST}{Richard B. Dunn Solar Telescope}   
\newacro{EB}{Ellerman bomb}
\newacro{EDP}{\'{E}dition Diffusion Presse Sciences}  
\newacro{EIT}{Extreme ultraviolet Imaging Telescope}
\newacro{EPIC}{European participation in Solar-C}
\newacro{ERC}{European Research Council}
\newacro{ESA}{European Space Agency}
\newacro{EST}{European Solar Telescope}
\newacro{EUV}{extreme ultraviolet}
\newacro{FAF}{flaring arch filament}
\newacro{FITS}{Flexible Image Transport System}
\newacro{FOV}{field of view}
\newacro{fov}{field of view}
\newacro{FWHM}{full width at half maximum}
\newacro{HAO}{High Altitude Observatory}
\newacro{HD}{hydrodynamics}
\newacro{Hi-C}{High Resolution Coronal Imager Sounding Rocket}
\newacro{HMI}{Helioseismic and Magnetic Imager}
\newacro{IAA}{Instituto de Astrof\'{i}sica de Andaluc\'{i}a}
\newacro{IAC}{Instituto de Astrof\'{i}sica de Canarias}
\newacro{IAS}{Institut d'Astrophysique Spatiale}
\newacro{IDL}{Interactive Data Language}
\newacro{IMaX}{Imaging Magnetograph eXperiment}
\newacro{INAF}{Istituto Nazionale di Astrofisica}
\newacro{IB}{IRIS bomb}
\newacro{IR}{infrared}
\newacro{IRIS}{Interface Region Imaging Spectrograph}
\newacro{ISAS}{Institute of Space and Astronautical Science}
\newacro{ISP}{Institute for Solar Physics}
\newacro{ISS}{International Space Station}
\newacro{ISSI}{International Space Science Institute}
\newacro{ITA}{Institute for Theoretical Astrophysics}
\newacro{JAXA}{Japan Aerospace Exploration Agency}
\newacro{KIS}{Kiepenheuer--Institut f\"{u}r Sonnenphysik}
\newacro{KPNO}{Kitt Peak National Observatory}
\newacro{LASP}{Laboratory for Atmospheric and Space Physics}
\newacro{LC}{liquid cristal}
\newacro{LMSAL}{Lockheed Martin Solar and Astrophysics Labratory}
\newacro{LOS}{line of sight}
\newacro{LTE}{local thermodynamic equilibrium}
\newacro{MC}{magnetic concentration}
\newacro{MCAO}{multi-conjugate adaptive optics} 
\newacro{MDI}{Michelson Doppler Imager}
\newacro{ME}{Milne-Eddington} 
\newacro{MHD}{magnetohydrodynamics}
\newacro{MOMFBD}{Multi-Object Multi-Frame Blind Deconvolution}
\newacro{MPE}{Max--Planck--Institut f\"ur extraterrestrische Physik}
\newacro{MPG}{Max--Planck--Gesellschaft}
\newacro{MPS}{Max Planck Institute for Solar System Research}
\newacro{MSSL}{Mullard Space Science Laboratory}
\newacro{MTF}{modulation transfer function}
\newacro{NAOJ}{National Astronomical Observatory of Japan}
\newacro{NASA}{National Aeronautics and Space Administration}
\newacro{NLTE}{non-local thermodynamic equilibrium}
\newacro{NOAA}{National Oceanic and Atmospheric Administration}
\newacro{non-E}{non-equilibrium}
\newacro{NSO}{National Solar Observatory}
\newacro{NWO}{Netherlands Organisation for Scientific Research}
\newacro{PRD}{partial redistribution}
\newacro{PROBA2}{PRoject for OnBoard Autonomy}
\newacro{PSF}{point spread function}
\newacro{QS}{quiet Sun}
\newacro{RAL}{Rutherford Appleton Laboratory}
\newacro{R-MHD}{radiation hydrodynamics}
\newacro{rms}{root mean square}
\newacro{RMS}{root mean square}
\newacro{ROB}{Royal Observatory of Belgium}
\newacro{ROI}{region of interest}
\newacro{RTE}{radiative transfer equation}
\newacro{SE}{statistical equilibrium}
\newacro{SDO}{Solar Dynamics Observatory}
\newacro{SJI}{slit-jaw image}
\newacro{SNR}{signal-to-noise ratio}
\newacro{SO}{Solar Orbiter}
\newacro{SoHO}{Solar and Heliospheric Observatory}
\newacro{SP}{Spectropolarimeter}
\newacro{SST}{Swedish 1-m Solar Telescope}
\newacro{SUMER}{Solar Ultraviolet Measurements of Emitted Radiation}
\newacro{SUFI}{Sunrise Filter Imager}
\newacro{SVD}{singular value decomposition}
\newacro{SVST}{Swedish Vacuum Solar Telescope}
\newacro{THEMIS}{T\'{e}lescope H\'{e}liographique pour l'Etude du 
   Magn\'{e}tisme et des Instabilit\'{e} Solaires}     
\newacro{TR}{transition region}
\newacro{TRACE}{Transition Region and Coronal Explorer}
\newacro{TSI}{total solar irradiance}
\newacro{UV}{ultraviolet}
\newacro{VIRGO}{Variability of solar IRradiance and Gravity Oscillations}
\newacro{VTT}{Vacuum Tower Telescope}    
\newacro{XRT}{X-Ray Telescope}

\def\acp#1{\pdftooltip{\acs{#1}}{\acl{#1}}}  

\long\def\startignore #1\stopignore{}   

\def\rmit#1{{\it #1}}              
           
\def\ie{\rmit{i.e.,}}              
\def\eg{\rmit{e.g.,}}              

\def\specchar#1{\uppercase{#1}}    
\def\specand{\,\&\,}               
\def\specand{ and }                

\def\CII{\mbox{C\,\specchar{ii}}} 
 
\def\CIV{\mbox{C\,\specchar{iv}}}

\def\CaII{\mbox{Ca\,\specchar{ii}}}

\def\FeI{\mbox{Fe\,\specchar{i}}} 
\def\FeII{\mbox{Fe\,\specchar{ii}}}

\def\MgII{\mbox{Mg\,\specchar{ii}}}

\def\MnI{\mbox{Mn\,\specchar{i}}} 
\def\MnII{\mbox{Mn\,\specchar{ii}}}

\def\NiII{\mbox{Ni\,\specchar{ii}}}

\def\OIV{\mbox{O\,\specchar{iv}}}

\def\SI{\mbox{S\,\specchar{i}}}

\def\SiIV{\mbox{Si\,\specchar{iv}}}

\def\Halpha{\mbox{H\hspace{0.1ex}$\alpha$}} 

\def\CaIIHK{\mbox{Ca\,\specchar{ii}\,\,H{\specand}K}}

\def\MgIIk{\mbox{Mg\,\specchar{ii}\,\,k}}
\def\MgIIh{\mbox{Mg\,\specchar{ii}\,\,h}}
\def\MgIIhk{\mbox{Mg\,\specchar{ii}{\specand}k}}
\def\hk{\mbox{h{\specand}k}}

\def\level #1 #2#3#4{$#1 \; ^{#2} \mbox{#3} ^{#4}$}   

\def\deg{\hbox{$^\circ$}}       

\def\arcsec{\hbox{$^{\prime\prime}$}}

\def\kms{\hbox{km$\;$s$^{-1}$}}

\def\is{\!=\!}                             
\def\={\hbox{$\!=\!$}}                     

\def\specchar#1{{\sc{#1}}}    
\def\rmit#1{#1}               

\def\revpar{}                 
\long\def\rev#1{#1}           

\def\deffigs{}

\newcommand{\gregalemail}{g.j.m.vissers@astro.uio.no}

\begin{document}

\title{Ellerman bombs at high resolution III.
Simultaneous observations with IRIS and SST}

\author{G. J. M. Vissers$^1$}
\author{L. H. M. Rouppe van der Voort$^1$}
\author{R. J. Rutten$^{2,1}$}
\author{M. Carlsson$^1$} 
\author{B. De Pontieu$^{3,1}$} 
\affil{${}^1$Institute of Theoretical Astrophysics, University of Oslo, 
  P.O. Box 1029 Blindern, N-0315 Oslo, Norway; 
\gregalemail}
\affil{${}^2$Lingezicht Astrophysics, 't Oosteneind 9, 4158CA Deil, 
  The Netherlands} 
\affil{${}^3$Lockheed Martin Solar and Astrophysics Laboratory, 
  3251 Hanover Street, Org.\ A021S, Bldg.\ 252, Palo Alto, CA\,94304,
  USA}

\shorttitle{Ellerman bombs with IRIS and SST}
\shortauthors{Vissers et al.}

\begin{abstract} 
Ellerman bombs are transient brightenings of the extended wings of the
solar Balmer lines in emerging active regions.
\revpar
We describe their properties in the ultraviolet lines sampled by
the Interface Region Imaging Spectrograph (IRIS), using simultaneous
imaging spectroscopy in \Halpha\ with the Swedish 1-m Solar Telescope
(SST) and ultraviolet images from the Solar Dynamics Observatory for
Ellerman bomb detection and identification.
We select multiple co-observed Ellerman bombs for detailed analysis.
The IRIS spectra strengthen  
\rev{the view that} Ellerman bombs mark reconnection between bipolar
kilogauss fluxtubes \rev{with} the reconnection and the resulting
bi-directional jet located within the solar photosphere and shielded
by overlying chromospheric fibrils in the cores of strong lines.
\rev{The spectra suggest} that the reconnecting photospheric gas
underneath is heated sufficiently to momentarily reach stages of
ionization normally assigned to the transition region and the corona. 
We also analyze similar outburst phenomena that we classify as small
flaring arch filaments and ascribe to higher-located reconnection.
They have different morphology and produce hot arches in
million-Kelvin diagnostics.
\end{abstract}


\section{Introduction}\label{sec:introduction}
\citetads{1917ApJ....46..298E} 
discovered intense short-lived brightenings of the extended wings of
the Balmer \Halpha\ line at 6563\,\AA\ that he called ``solar hydrogen
bombs''.
They are called Ellerman bombs (henceforth EB) since
\citetads{1960PNAS...46..165M}. 
For more detail we refer to the excellent summary by
\citetads{2002ApJ...575..506G} 
and our more recent review of the extensive \acp{EB} literature in
\citetads{2013JPhCS.440a2007R}. 

We discuss the subsequent \acp{EB} literature below, but here point
out the recent discovery by
\citetads{2014Sci...346C.315P} 
of very hot ``bombs'' in ultraviolet spectra from the Interface Region
Imaging Spectrograph (IRIS,
\citeads{2014SoPh..289.2733D}). 
\rev{The present paper addresses their \rev{suggestion}
that these bombs might have been \acp{EB}s or similar
to \acp{EB}s.}

A major motivation to study \acp{EB}s is that they supposedly mark
locations of serpentine flux rope emergence in newly emerging active
regions (\eg\
\citeads{2002SoPh..209..119B}; 
\citeads{2004ApJ...614.1099P}; 
\citeads{2007ApJ...657L..53I}; 
\citeads{2009A&A...508.1469A}; 
\citeads{2009ApJ...701.1911P}). 
Understanding their nature may therefore present a way to measure
active region evolution, in particular the reconnective field
topography evolution that eventually produces much larger solar
outbursts.
In this context, \acp{EB}s should become useful as telltales of
strong-field reconnection when well understood.

In addition, the complex physics and spectrum formation of the
\acp{EB} phenomenon are of interest per s\'e since \acp{EB}s appear to
be pockets of hot gas within the photosphere.
The discovery of extremely hot \acp{IRIS} bombs by
\citetads{2014Sci...346C.315P} 
that also appear to be photospheric enhances this interest.
In our present series of \acp{EB} analyses we employ high-quality
imaging spectroscopy with the \acl{SST} (SST;
\citeads{2003SPIE.4853..341S}) 
to study \acp{EB}s at unprecedented spatial, spectral, and temporal
resolution.
Paper~I (\citeads{2011ApJ...736...71W}) 
established that \acp{EB}s are a purely photospheric phenomenon.

Paper~II (\citeads{2013ApJ...774...32V}) 
added evidence that \acp{EB}s mark magnetic reconnection of strong
opposite-polarity field concentrations in the low photosphere and
discussed their appearance in 1700\,\AA\ images from the \acl{AIA}
(AIA; \citeads{2012SoPh..275...17L}) 
of the \acdef{SDO}.

\bibnote{2011ApJ...736...71W}{(Paper~I)}
\def\PaperI{\href{http://adsabs.harvard.edu/abs/2011ApJ...736...71W}
{Paper~I}}

\bibnote{2013ApJ...774...32V}{(Paper~II)}
\def\PaperII{\href{http://adsabs.harvard.edu/abs/2013ApJ...774...32V}
{Paper~II}}

Let us morphologically define the three bomb-like phenomena that we
discuss below, based on our inspections of dozens of such features in
\acp{SST}, \acp{SDO}, and \acp{IRIS} data.
More detail on their recognition is given in
Sect.~\ref{sec:observations}.

We define ``Ellerman bombs'' (EB) as substantial brightenings of the
extended wings of \Halpha\ without core brightening that, at
sufficient angular and temporal resolution, show definite rapid-flame
morphology when viewed from aside as described in \PaperI.
\acp{EB} \Halpha\ wing brightenings exceed those from much more
ubiquitous magnetic concentrations that happen to also appear bright
in the \Halpha\ wings (``pseudo-EBs'',
\citeads{2013JPhCS.440a2007R}). 

Next, we define ``flaring arch filaments'' (henceforth FAFs) as sudden
fierce brightenings in \acp{AIA} 1600\,\AA\ image sequences that
differ from the \acp{EB} brightenings also seen in this \acp{AIA}
channel by appearing with shorter duration and more abrupt changes,
having elongated morphology, and showing fast apparent brightness
motion along filamentary strands.
Because they are usually much less evident in \acp{AIA} 1700\,\AA\
images, their 1600\,\AA\ appearance is likely due to brightening of
the \CIV\ doublet at 1548 and 1550\,\AA\ in \acp{AIA}'s 1600\,\AA\
passband.
Their filamentary morphology and rapid evolution suggest that these
are heating events, likely reconnection, that take place along the
fibrilar canopy seen \eg\ at \Halpha\ line center, or eject heated
matter along chromospheric field lines.

Finally, we define ``IRIS bombs'' (henceforth IBs) \rev{following
\citetads{2014Sci...346C.315P}} 
as ultraviolet brightenings with substantial emission in the \SiIV\
lines observed by \acp{IRIS}, and showing these with very wide and
complex non-Gaussian profiles on which deep absorption blends of lower
metal ionization stages are superimposed.

The visibility of \acp{EB}s in \acp{AIA}'s 1700\,\AA\ images and the
wonderful full-time full-disk availability of \acp{SDO} data enables
one to check any new observational \acp{EB} study to ascertain that it
does not instead address pseudo-EBs or \acp{FAF}s.
We did so for the \acp{EB} literature since \PaperI\ and briefly
comment on our scrutiny here.

We judge that the 3570 \acp{EB}s of
\citetads{2013SoPh..283..307N} 
were probably pseudo-EBs, and likewise the 4 \acp{EB}s of
\citetads{2013ApJ...779..125N}. 
Both studies targeted decaying sunspots rather than emerging active
regions.

In contrast, we recognized the 3 \acp{EB}s of
\citetads{2013A&A...557A.102B} 
as well-defined \acp{EB}s in 1700\,\AA, indeed occurring in a complex
region with much flux emergence.
Similarly for at least EB3 and EB4 of
\citetads{2014ApJ...792...13H} 
and the single \acp{EB} of
\citeads{2013SoPh..288...39Y} 
(which occurred a day after those of
\citeads{2013A&A...557A.102B} 
in the same region).
Most recently, the near-limb \acp{EB}s in
\citetads{2015ApJ...798...19N} 
have obvious flame morphology.

Generally, the latter papers confirm our view in \PaperI, \PaperII,
\citetads{2013JPhCS.440a2007R}. 
\acp{EB}s are strong-field opposite-polarity cancelations that occur
in complex emerging active regions.
They mark reconnection taking place in the photosphere, and produce
substantial local heating that leaves no direct signature in the
overlying chromosphere and transition region.

Modeling of the \Halpha-wing enhancements that characterize \acp{EB}s
was recently reported by
\citetads{2013ApJ...779..125N}, 
\citetads{2013A&A...557A.102B}, 
\citetads{2014ApJ...792...13H}, 
and \citetads{2014A&A...567A.110B}. 
We return to these analyses in Sect.~\ref{sec:discussion}, but already
point out here that they agree with all earlier modeling in claiming
upper-photosphere temperature enhancements of only 1000--5000\,K,
in obvious conflict with the notion that \acp{EB}s might be \acp{IB}s
for which \citetads{2014Sci...346C.315P} 
suggest formation temperatures near 100\,000~K.

Were the \acp{IB}s of
\citetads{2014Sci...346C.315P} 
indeed \acp{EB}s \rev{as suggested by them}?  
Our similar inspection of the concurrent \acp{AIA} 1700 and 1600\,\AA\
morphology turned out indecisive.
Their bomb B-1, with deep \NiII\ and \FeII\ absorption blends in the
\SiIV\ lines, seemed a bonafide \acp{EB} to us but the others looked
more \acp{FAF}-like.
Hence, as stressed by
\citetads{2014Sci...346C.315P}, 
there is a clear need for simultaneous \acp{IRIS} and ground-based
\Halpha\ observation of \acp{EB}s and \acp{IB}s because \acp{EB}
recognition is easier in \Halpha.

In this paper we address this \acp{EB}--\acp{IB} issue by combining
new \acp{EB} and \acp{FAF} observations with the \acp{SST} with
simultaneous observations with \acp{IRIS} and \acp{SDO}/\acp{AIA}.
Our conclusion is that both \acp{EB}s and \acp{FAF}s produce
ultraviolet line profiles of \acp{IB} type, and that these provide
valuable insights and constraints.

The observations are presented in the next section, the results in
Sect.~\ref{sec:results}.
We discuss them in Sect.~\ref{sec:discussion} and add conclusions in
Sect.~\ref{sec:conclusions}.

\begin{table*}[bth]
\caption{Overview of the datasets analyzed in this study.}
\begin{center}
\begin{tabular}{llccclllcccrc}
        \hline \hline
  {}      & {}    & \multicolumn{3}{c}{Target}                & {} & {} & \multicolumn{4}{c}{Diagnostic details} & {} & {} \\ 
  \cline{3-5} \cline{8-11}
  Set & Date  & AR  & ($X$,\,$Y$) & $\theta$ & Instru- & OBSID &
  Name & $\lambda_{0}$ & Range$^{a}$ & $\Delta \lambda$ & $\Delta t$ & Time \\
  {} & {}  & {}  & [\arcsec] & [\deg] & ment & {} & {} & [\AA] & [\AA] & [m\AA] & [s] & [UTC] \\
  \hline
  1 & 2013 Sep 6 & 11836 & (763,\,129) & 50.6 & CRISP & --- & \Halpha & 6563 & $\pm$1.2 & 100 & 5.5 & 08:15\,--\,09:01 \\
  {} & {} & {} & {}  & {} & IRIS & 4003004168 & SG (4$\times$1\arcsec) & ---  &
  ---& --- & 11 & 08:11\,--\,11:39 \\
  {} & {} & {} & {}  & {} & {} & {} & SJI 1330 & 1340 & 55  & --- & 12 & {} \\
  {} & {} & {} & {}  & {} & {} & {} & SJI 1400 & 1390 & 55  & --- & 12 & {} \\
  {} & {} & {} & {}  & {} & {} & {} & SJI 2796 & 2796 & 4  & --- & 12 & {} \\
  {} & {} & {} & {}  & {} & {} & {} & SJI 2832 & 2830 & 4  & --- & 69 & {} \\
  \hline
  2 & 2014 Jun 14 & 12089 & (221,\,278) & 21.5 & CRISP & --- & \Halpha & 6563 & $\pm$1.4 & 200 & 11.4 & 07:20\,--\,08:11 \\
  {} & {} & {} & {}  & {} & {} & --- & \CaII & 8542 & $\pm$1.2 & 100 & {} & {} \\
  {} & {} & {} & {}  & {} & {} & --- & \FeI & 6302 & $-$0.048$^{b}$ & --- & {} & {} \\
  {} & {} & {} & {}  & {} & IRIS & 3820256197 & SG (96$\times$0.33\arcsec) & ---
  & --- & --- & 516 & 07:29\,--\,10:47 \\
  {} & {} & {} & {}  & {} & {} & {} & SJI 1330 & 1340 & 55  & --- & 17 & {} \\
  {} & {} & {} & {}  & {} & {} & {} & SJI 1400 & 1390 & 55  & --- & 17 & {} \\
  {} & {} & {} & {}  & {} & {} & {} & SJI 2796 & 2796 & 4  & --- & 17 & {} \\
  {} & {} & {} & {}  & {} & {} & {} & SJI 2832 & 2830 & 4  & --- & 86 & {} \\
  \hline
  3 & 2014 Jun 15 & 12089 & (411,\,281) & 31.0 & CRISP & --- & \Halpha & 6563 & $\pm$1.4 & 200 & 11.4 & 07:47\,--\,08:49 \\
  {} & {} & {} & {}  & {} & {} & --- & \CaII & 8542 & $\pm$1.2 & 100 & {} & {} \\
  {} & {} & {} & {}  & {} & {} & --- & \FeI & 6302 & $-$0.048$^{b}$ & --- & {} & {} \\
  {} & {} & {} & {}  & {} & IRIS & 3820256197 & SG (96$\times$0.33\arcsec) & ---
  & --- & --- & 516 & 07:29\,--\,10:55 \\
  {} & {} & {} & {}  & {} & {} & {} & SJI 1330 & 1340 & 55  & --- & 17 & {} \\
  {} & {} & {} & {}  & {} & {} & {} & SJI 1400 & 1390 & 55  & --- & 17 & {} \\
  {} & {} & {} & {}  & {} & {} & {} & SJI 2796 & 2796 & 4  & --- & 17 & {} \\
  {} & {} & {} & {}  & {} & {} & {} & SJI 2832 & 2830 & 4  & --- & 86 & {} \\
  \hline \hline
\end{tabular}
\begin{minipage}{.9\hsize}
  $^{a}$ This column gives the passband width in case of the slitjaw image channels. \newline
  $^{b}$ Full Stokes polarization measurements were obtained at this
  wavelength position.
\end{minipage} 
\end{center}
\label{tab:datasets}
\end{table*}

\begin{figure*}
  \centerline{\includegraphics[width=\textwidth]{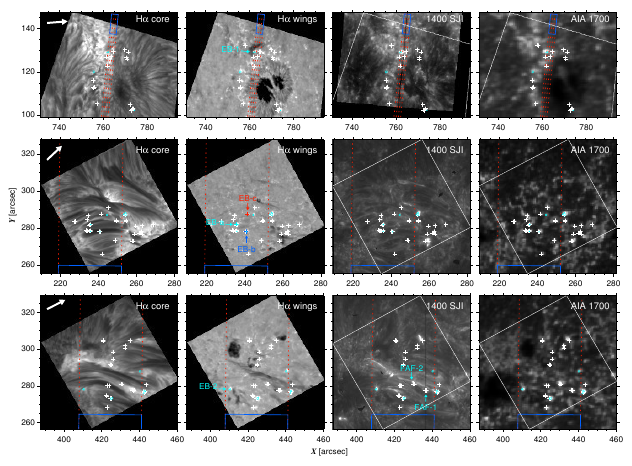}}
  \caption[]{\label{fig:fovs} 
  Field-of-view images from datasets 1--3 \rev{in rows from top to
  bottom}.
  The \acp{SST} and \acp{IRIS} images have been rotated to solar
  $(X,Y)$ coordinates.
  {\it Left to right:\/} \acp{SST} \Halpha\ core, \acp{SST} \Halpha\
  summed wings, \acp{IRIS} 1400\,\AA\ slitjaw (SJI = slitjaw image),
  \acp{AIA} 1700\,\AA.
  Red dashed lines specify \acp{IRIS} slit locations ({\it top row\/})
  or raster extent ({\it lower rows\/}).
  \Halpha\ \acp{EB} detections for the instant sampled by these images
  are marked with cyan contours.
  White pluses mark \Halpha\ \acp{EB} detections at other times in
  each data sequence.
  The selected \acp{EB}s presented in detail below are identified in
  the second column: EB-1 in the top row, EB-2 in the bottom
  row, and ``weak'' EB-a, EB-b, and EB-c in the middle row. 
  The position of the selected \acp{FAF}s (FAF-1 and FAF-2) are
  indicated in the third panel of the bottom row.
  The white arrows in the first column specify the direction to the
  nearest limb.
  The white frames on the \acp{IRIS} and \acp{AIA} panels specify the
  \acp{SST} field of view (the actual \acp{IRIS} slitjaw images are
  larger than shown here).
  The blue rectangles outline the quieter areas over which \acp{IRIS}
  spectra were averaged to obtain reference profiles.
  In the lower rows these extend 5.6 and 10.3~arcsec below the image
  cutouts, respectively.
  Animated versions of these these panels, including \acp{SDO}/\acp{HMI}
  line-of-sight magnetograms, are available in the online edition of the
  journal.
  }
\end{figure*}

\section{Observations, Reduction, Methods}\label{sec:observations}

\subsection{Data collection}
For this study we analyzed data from multiple observing campaigns
targeting emerging active regions with co-pointing of the \acp{SST}
and \acp{IRIS}.
The \acp{SST}'s \acl{CRISP} (CRISP;
\citeads{2008ApJ...689L..69S}), 
a Fabry-P\'erot interferometer, collected imaging spectroscopy in at
least \Halpha\ \rev{(further specification in Table~\ref{tab:datasets}).}

\rev{The \acp{IRIS} spectrograph (SG)} recorded its standard selection
of ultraviolet lines: the \CII\ doublet near 1335\,\AA, the \SiIV\
doublet at 1394\,\AA\ and 1403\,\AA, the \MgIIhk\ lines near
2796\,\AA\ and 2804\,\AA\ including the overlapping-wing part between
them with various blends, in particular the \MgII\ triplet lines near
2798\,\AA\ (which are two overlapping transitions that overlap so
closely that they look like one line in the spectrum).
For more details including characteristic formation temperatures see
Table~4 of \citetads{2014SoPh..289.2733D}. 
For \acp{IRIS}-related formation studies of \MgIIhk\ see
\citetads{2013ApJ...772...89L}, 
\citetads{2013ApJ...772...90L}, 
and
\citetads{2013ApJ...778..143P}; 
see
\citetads{2015arXiv150401733P} 
for a similar formation study of the \MgII\ triplet lines.
In addition, \acp{IRIS} collected slitjaw images (SJI) \revpar\rev{in the
1330\,\AA, 1400\,\AA, 2796\,\AA\ and 2832\,\AA\ channels}, 
as detailed in Table~3 of
\citetads{2014SoPh..289.2733D}. 

\rev{In Table~\ref{tab:datasets} we specify pointing, spectral and timing
details} for the three particular datasets from these co-ordinated
observing campaigns \rev{that were selected for} this paper.

The top row of Fig.~\ref{fig:fovs} shows image samples of the first.
It covered the major sunspot in active region AR\,11836 that had a
pronounced moat flow.
\revpar
\acp{IRIS} supported these observations 
\revpar
in a ``4-step sparse raster'' mode of slit motion, covering 3~arcsec
\revpar 
with 2~s exposure times \rev{per step}. 
\revpar
This pattern gives good temporal resolution for a given spot on the
Sun but smaller chance of hitting a scarce feature such as an
\acp{EB}.~ 
\revpar

\revpar 
Sample images for datasets 2 and 3 \rev{covering AR\,12089} are similarly shown in the
lower rows of Fig.~\ref{fig:fovs}.
\revpar
\rev{For both \acp{IRIS} took} a dense synoptic raster of
96 steps\rev{, covering 31.35~arcsec} at \revpar
4\,s exposure time \rev{per step}.
\revpar
The larger pattern width gives larger spatial chance of hitting an
\acp{EB}, but the \rev{consequently} low repeat cadence, longer than the typical 
\acp{EB} appearance, diminishes the catch.
\revpar

We also collected corresponding image sequences from
\acp{SDO}/\acp{AIA} and \acp{SDO}/\acp{HMI}
(\citeads{2012SoPh..275..207S}) 
using the JSOC image cutout service at Stanford University.

\subsection{Data reduction}
The \acp{SST}/\acp{CRISP} data were reduced using the CRISPRED
pipeline (\citeads{2015A&A...573A..40D}). 
It includes (1) dark and flat field correction, (2) multi-object
multi-frame blind deconvolution
(\citeads{2005SoPh..228..191V}) 
to reduce the effects of high-order atmospheric seeing, (3)
minimization of remaining small-scale deformation through
cross-correlation
(\citeads{2012A&A...548A.114H}), 
(4) prefilter transmission correction
(\citeads{2010PhDT.......219D}), 
(5) correction for time-dependent image rotation due to the
alt-azimuth telescope configuration, and (6) removal of remaining
rubber-sheet distortions by destretching
(\citeads{1994ApJ...430..413S}). 

The \acp{SST} and \acp{IRIS} data were co-aligned using far-wing
images in \Halpha\ (dataset 1) or \CaII~8542\,\AA\ (datasets 2 and 3)
and the \acp{IRIS} \MgIIh~2832\,\AA\ slitjaws.
\acp{SDO}/\acp{AIA} 1600\,\AA\ or 1700\,\AA\ images (depending on the
dataset) were used as initial co-location reference to define common
features in the fields-of-view and their offsets.
The \acp{SST} data were then resampled to the \acp{IRIS} slitjaw pixel
size of 0.167~arcsec$^2$. 
Finally, sub-images (usually containing one or more pores) were then
selected manually for cross-correlation at each time step.

The \acp{AIA} and \acp{HMI} image sequences were also precisely
co-aligned with the full-resolution \acp{SST} image sequences.
\revpar

In the alignment and the data analysis we made much use of the
\acl{CRISPEX} (CRISPEX;
\citeads{2012ApJ...750...22V}) 
for data browsing. 
The latest version (available through SolarSoft) can handle both
\acp{FITS}-formatted \acp{IRIS} data including slitjaw images, and
legacy ``La Palma''-format data files from the \acp{SST}.

\subsection{\acp{EB} identification using \Halpha}
Identifying \acp{EB}s is not a trivial matter.
A substantial part of the \acp{EB} literature did not address actual
\acp{EB}s but ``pseudo-EBs'': magnetic concentrations (MC) in network
or plage that likewise brighten in the \Halpha\ wings as explained by
\citetads{2006A&A...449.1209L}. 
Such \acp{MC} brightening is more familiar as ``facular bright
points'' in the continuum and in the molecular G-band around
4305\,\AA\ and as ``line gaps'' in neutral-metal lines, but it
actually reaches largest contrast in the blue wing of \Halpha\
(\citeads{2006A&A...452L..15L}). 
It is well understood and is not a sign of heating but of
deeper-than-normal radiation escape (summary and references in
\citeads{2013JPhCS.440a2007R}). 
Hence, care must be taken to ascertain that features that appear
bright in an \Halpha\ wing are indeed \acp{EB}s and not just facular
brightenings---a warning already given by
\citetads{1917ApJ....46..298E} 
himself.  
A first check is to ascertain on the daily magnetogram movies from
\acp{SDO}/\acp{HMI} that the observed field of view is part of an
active region with much flux emergence and fast streaming motions
including bipolar collisions.

In \PaperI\ and \PaperII\ we found that \acp{EB}s are best identified
using \Halpha\ wing images with slanted viewing away from disk center.  
We obtain such wing images by summing the three spectral samplings of
both \Halpha\ wings around $\Delta \lambda = \pm 1.0$\,\AA.
In limbward viewing at the \acp{SST} resolution \acp{MC}s reach less
\Halpha-wing brightness contrast than near disk center, whereas
\acp{EB}s appear with definite flame morphology.
They show up as elongated bright upright features that rapidly flicker
(hence ``flames'') during a few minutes while their feet are anchored
in and travel along \acp{MC}-rich intergranular lanes.
Their tops extend intermittently up to Mm heights.
This rapid-flame behavior is the best diagnostic to classify an
\Halpha\ wing brightening as \acp{EB}\footnote{We emphatically invite
the reader to inspect the high-resolution \acp{EB} movies available
with \PaperI\ and so become familiar with this defining morphology.}.

In \PaperI\ this flame morphology was used to identify \acp{EB}s
manually, but in \PaperII\ we defined automated selection criteria
employing the brightness contrast, spatial extent, temporal continuity
and lifetime of candidate features in \acp{SST} imaging-spectroscopy
sequences sampling the \Halpha\ wings.
These criteria were tuned to optimally recover \rev{example} \acp{EB}s
that had been identified visually from their time-dependent morphology,
\rev{and then applied to obtain a faster, more objective, and more complete
identification of \acp{EB}s in each data set}.

In the present analysis we have applied these criteria to our three
datasets, but with an adjustment for dataset 2 in which we lowered the
thresholds to 145\% brightness contrast over the field-of-view mean
for the \acp{EB} kernel and 130\% for adjacent pixels, instead of the
\PaperII\ values of 155\% and 140\%.
We did so because \rev{visual inspection showed that} with the latter
thresholds we missed a number of features of which the morphology
suggested they were \acp{EB}s, even though weak in relative
\Halpha\ wing excess.
A reason may be that this field of view contained no dark umbrae or
pores and therefore had a higher mean profile than in \PaperII.
Also, it was the closest to disk center where \acp{MC}s appear
brightest in the \Halpha\ wings.

Examples of the resulting threshold contours outlining \acp{EB}
candidates are shown in Fig.~\ref{fig:fovs} for those that were
detected at the particular moment at which each image was taken.
The numerous white plus signs mark other \Halpha\ \acp{EB} detections
during the whole \acp{SST} observing period.
Some overlap closely and are detections of repetitive \acp{EB}s at
about the same location.

\subsection{\acp{EB} identification in the 1700\,\AA\ continuum}
In \PaperII\ we also tried to automate \acp{EB}-finding in
\acp{SDO}/\acp{AIA} 1700\,\AA\ image sequences.  
\acp{EB}s typically appear in these as strongly enhanced, fairly
pointlike and fairly stable brightness features. 
\acp{AIA} 1600\,\AA\ images show them at yet larger brightness
contrast above ordinary \acp{MC}s, but as noted above the scene at
this wavelength often contains \acp{FAF}s as well
(\citeads{2013JPhCS.440a2007R}). 

In the less \acp{FAF}-infested 1700\,\AA\ images a contrast criterion
of 8\,$\sigma$ above the mean intensity was found to recover most of
the brighter \Halpha\ \acp{EB}s.
This conservative threshold may miss weaker \acp{EB}s, but lower
cutoff values give more confusion with non-eruptive \acp{MC}s.
An additional lifetime maximum of 5~min was also set to distinguish
\acp{EB}s from longer-lived \acp{MC}s. 
A further non-automated check is to ignore 1700\,\AA\ detections when
they exhibit \acp{FAF} behavior at 1600\,\AA.

\subsection{\acp{EB} visibility in \acp{IRIS} slitjaw images and spectra}
Blinking of the \acp{IRIS} slitjaw and the \acp{CRISP} \Halpha\ movies
suggested that bright \acp{EB}s detected in the \Halpha\ wings often
show up as bright features in the \CII, \SiIV, and \MgIIk\ slitjaw
images.
However, the fraction for which we also have \acp{IRIS} spectra is
small.
For dataset 1 this is obvious in the top row of Fig.~\ref{fig:fovs}
where the narrow \acp{IRIS} scan-strip missed most \Halpha\ \acp{EB}
detections (white plus signs).

For datasets 2 and 3 the \acp{IRIS} raster extent was much wider so
that many more \Halpha\ \acp{EB}s fell within it, but the slow raster
repeat cadence of 516\,s meant that most of these were not sampled
spectroscopically during their brief lifetime.  
Nevertheless, in the few cases of proper \acp{EB} slit coverage the
\acp{IRIS} spectra show corresponding brightening of the major
\acp{IRIS} lines, so that it is probable that the slitjaw brightenings
are simply set by the wavelength-integrated enhancements of these
lines.

Examples are shown in Fig.~\ref{fig:fovs} and yet clearer in the
smaller cutout sequences in Figs.~\ref{fig:cutouts_EB1},
\ref{fig:cutouts_EB2}, and \ref{fig:cutouts_EBabc} discussed below.
In general, the bright \acp{IRIS} slitjaw features are not one-to-one
identical with the \Halpha-wing brightenings but there is good overall
correspondence in location, orientation, shape, and evolution.  

\subsection{\acp{EB} and \acp{FAF} selection for  presentation here}
For dataset 1 application of the \PaperII\ \Halpha\ criteria resulted
in 31 \Halpha\ \acp{EB} detections, of which many showed pronounced
slitjaw brightening.
However, of those 31 \Halpha\ \acp{EB}s only 4 where sampled by the
\acp{IRIS} slit and only one of these showed pronounced brightening in
the slitjaws.
We selected the latter one for detailed presentation below and
henceforth refer to this \acp{EB} as EB-1.
It also passed the 1700\,\AA\ criteria of \PaperII, while the other
three sampled by the \acp{IRIS} slits did not.

Data set 2 was closest to disk center so that distinction from
ordinary \acp{MC}s and recognition of flame morphology was likely
hampered by top-down viewing.
Our down-tuned \Halpha\ criteria gave 49 \acp{EB} detections, of which
6 were sampled spectroscopically by \acp{IRIS}.
Most were weak in \Halpha\ and the \acp{IRIS} spectra, and weak or
invisible in the \acp{IRIS} slitjaws.  
We selected the three with the highest \acp{IRIS} profiles and call
them EB-a, EB-b, EB-c henceforth and present their spectra below as
examples of weaker or even questionable \acp{EB}s. 
Only EB-b passed the 1700\,\AA\ criteria.

For dataset~3 the \PaperII\ \Halpha\ criteria yielded 56 detections of
which 10 were sampled spectroscopically by \acp{IRIS}.
Most were weak; the exception was a very long-lived repetitive
\acp{EB} which we call EB-2 henceforth. 
It did not pass the \PaperII\ 1700\,\AA\ criteria initially, but it
did when we relaxed the constraint on lifetime to distinguish weak
\acp{EB}s from longer-lived \acp{MC}s.
EB-2 was clearly not a pseudo-EB in \Halpha. 
In 1700\,\AA\ it occurred repetitively for an exceptionally long
period of time.

In dataset~3 we also noted a string of fierce repetitive
interconnected brightenings of which the \acp{SDO}/\acp{AIA}
1600\,\AA\ movie shows they were \acp{FAF}s.
We also selected two of these for comparative display and discussion
below.

A fourth \acp{SST}-\acp{IRIS} dataset taken on September 25, 2013 was
discarded because of its 48 \Halpha\ \acp{EB} detections only one was
sampled by \acp{IRIS}.
It did not show up in the slitjaw images and produced only slight
ultraviolet line brightenings, much as the discarded \acp{EB}s in
dataset 2 and therefore, like those, not selected for detailed
presentation here.
There were more \acp{EB}s visible in the latter, but outside the
narrow raster strip.

\section{Results} \label{sec:results}

In this section we present the observations for each selected feature
in succession, using the same plot formats for cutout samples from the
\acp{SST}, \acp{SDO} and \acp{IRIS} slitjaw images, light curves
distilled from these, and \acp{IRIS} spectra at selected pixels and
times corresponding to the cutouts.
For each feature we add some interpretation, but we postpone overall
discussion to Sect.~\ref{sec:discussion}.

\subsection{Details for EB-1}

When viewing EB-1 in the \Halpha\ sequences, using \acp{CRISPEX} to
inspect its spatial, temporal, and spectral behavior in flexible
cursor-controlled movie mode (a recommended modus operandi), this
\acp{EB} appears as a sequence of unmistakable tall \acp{EB} flames in
the outer \Halpha\ wings, re-occurring in rapid succession during the
whole observing period, with fast motion of its successive footpoints
away from the spot along an \acp{MC}-filled lane.
Our \acp{CRISPEX} inspection also showed that a dark redshifted
chromospheric fibril was overlying the \Halpha\ core part of the time.

In the \acp{AIA} 1700\,\AA\ sequence EB-1 also stands out bright and
\acp{EB}-like, \ie\ pointlike, roughly stationary, without filamentary
\acp{FAF} signature. 
The \acp{HMI} magnetogram sequence shows that it occurred in a complex
region with much streaming motion, from the sunspot towards extended
plage and pores of both polarities further North.  
Small patches of opposite polarity traveled fast in this flow but were
barely visible with \acp{HMI}.
Likely such patches produced EB-1 successively wile canceling.
In \PaperII\ we observed \acp{EB}s at similar cancelations of small
opposite-polarity patches in \acp{SST} magnetograms with better detail
than given by \acp{HMI}.

Unfortunately, EB-1 was sampled by the \acp{IRIS} slit only at the
beginning of the \acp{SST} sequence and only at two positions of the
narrow scan pattern.
In its successive flarings EB-1 migrated eastward out of the scan
strip.
However, during this 10-min overlap period the rapid-scan format gave
good temporal sample resolution. 

\begin{figure}
  \centerline{\includegraphics[width=\columnwidth]{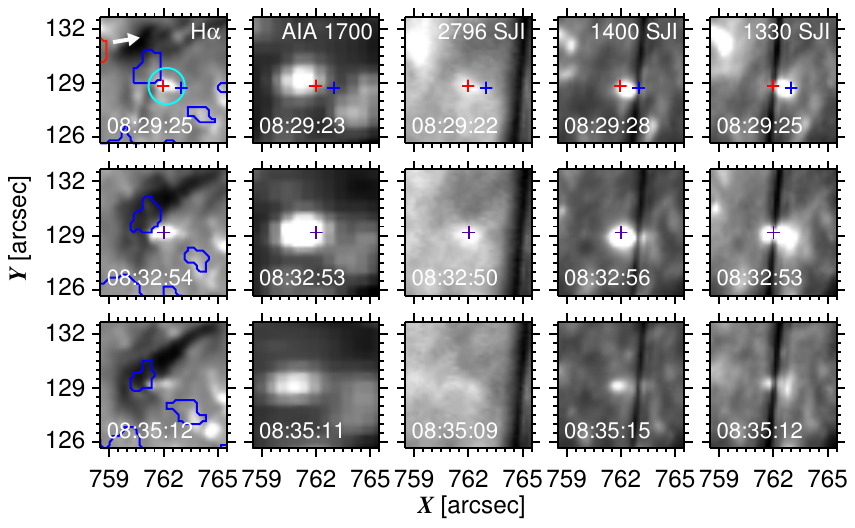}}
    \caption[]{\label{fig:cutouts_EB1} 
    Time evolution of EB-1 in various diagnostics.
    {\it Left to right:\/} co-spatial image cutouts of \acp{CRISP}
    \Halpha\ wings, \acp{AIA} 1700\,\AA, \acp{IRIS} \MgIIk, \SiIV, and
    \CII\ slitjaws.
    \rev{Red and blue contours in the first column indicate patches of
    positive and negative polarity (thresholded at 0 and -1000
    counts), respectively, based on the HMI line-of-sight magnetic
    field data.}
    The time of observation is specified at the bottom of each panel.
    The plus signs mark the locations for which corresponding spectra
    are shown with the same color coding in
    Fig.~\ref{fig:spectra_EB1}.
    \rev{The cyan circle indicates the size (and instantaneous
    location) of the mask used to determine the light curves in
    Fig.~\ref{fig:lightcurves_EB1}}
    The arrow in the \rev{same} panel specifies the ``upright'' direction
    to the nearest limb.
    The panels in each column have been bytescaled to the same
    extremes.
    }
\end{figure}

Figure~\ref{fig:cutouts_EB1} shows a selection of small-field cutouts
of EB-1 in various diagnostics.
The time differences along rows are negligible by using
nearest-neighbor selection, whereas the rows are about 3~min apart in
order to sample EB-1's evolution.
Colored plus signs specify the pixels for which \acp{IRIS} spectra are
shown in Fig.~\ref{fig:spectra_EB1} with the same colors.
The slitjaws in the last two columns of Fig.~\ref{fig:cutouts_EB1}
show the \acp{IRIS} slit as a dark near-vertical stripe at or close to
the plus signs at corresponding times.

\begin{figure}
\centerline{\includegraphics[width=\columnwidth]{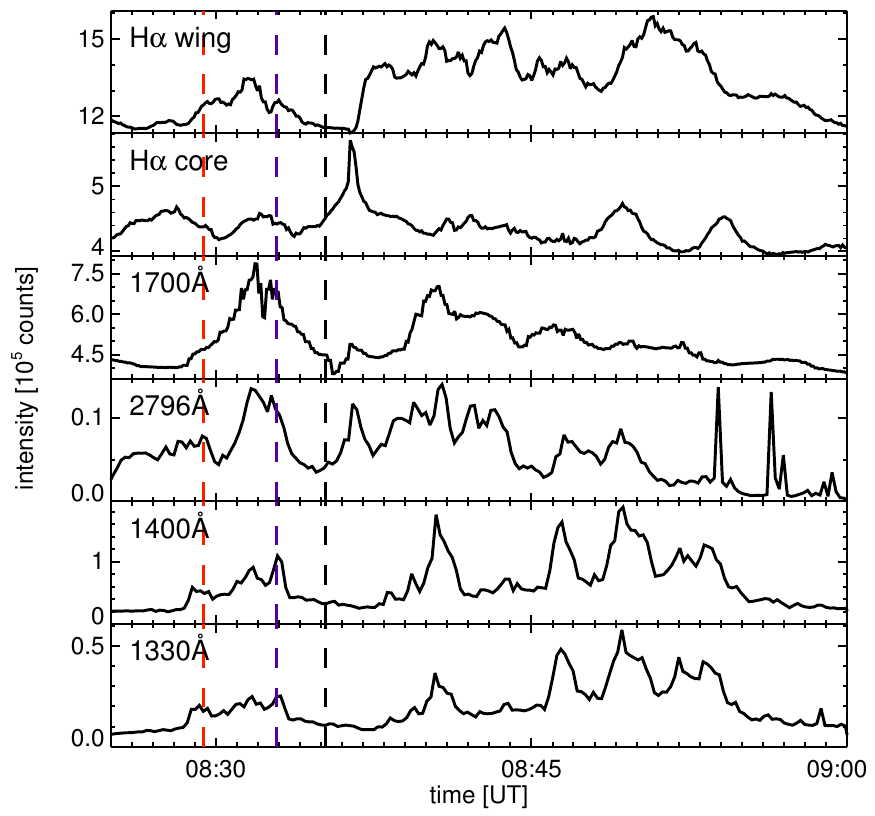}}
  \caption[]{\label{fig:lightcurves_EB1} 
  Light curves for EB-1 showing from top to bottom the intensities in
  the \Halpha\ wings and core, \acp{AIA} 1700\,\AA, and the \MgIIk,
  \SiIV\ and \CII\ slitjaws as function of time.
  Each intensity value is the sum over a co-moving circular aperture
  centered on the \Halpha\ detection with a diameter of 2.0~arcsec
  (1.5 times the maximum diameter of the \Halpha\ \acp{EB}
  contour)\rev{, as indicated by the circle in the first panel of
  Fig.~\ref{fig:cutouts_EB1}}.
  The presence of the slit and diffraction from it produced extra
  modulation of the lower three diagnostics during the first ten
  minutes. 
  The vertical dashed lines mark the times per row of
  Fig.~\ref{fig:cutouts_EB1}, with the same color coding.
  The first ({\it red\/}) corresponds to the near-simultaneous blue
  and red sampling of EB-1 in the first row, the second to the violet
  colored sampling in the second row, the third ({\it black\/}) to the
  \acp{EB} aftermath in the third row for which we show no \acp{IRIS}
  spectra (since not of interest).
  }
\end{figure}

\begin{figure*}
\centerline{\includegraphics[width=\textwidth]{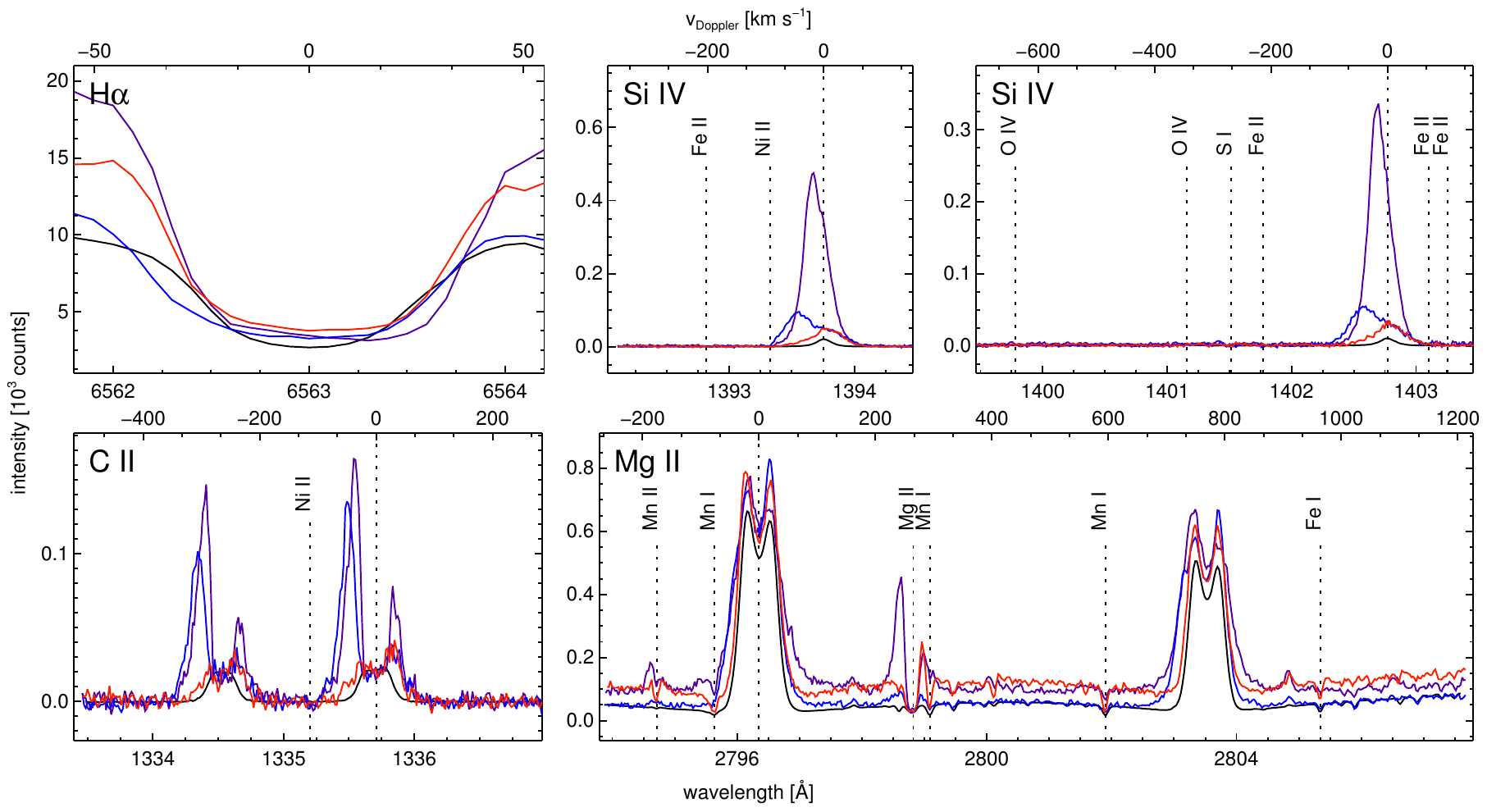}}
   \caption[]{\label{fig:spectra_EB1} 
   \acp{CRISP} and \acp{IRIS} spectra of EB-1.
   Clockwise: \Halpha, the \SiIV\ lines near 1394\,\AA\ and 1403\,\AA,
   the \MgIIhk\ lines near 2796\,\AA\ and 2804\,\AA\ with between them
   the (double) \MgII\ triplet lines near 2798\,\AA, and the \CII\
   doublet near 1335\,\AA.
   Several other lines and blends are indicated by labeled vertical
   dashed lines.
   All wavelengths are vacuum values (against the convention of
   specifying air wavelengths above 2000\,\AA\ because \acp{IRIS} is a
   space platform).
   The red profiles were taken at 08:29:25~UT, the blue profiles at
   08:29:28~UT, the violet profiles at 08:32:53~UT.
   The colors correspond to the pixel markers in
   Fig.~\ref{fig:cutouts_EB1}.
   The black profiles show the average spectrum for the ``quiet-Sun''
   reference box in the first row of Fig.~\ref{fig:fovs}.
   Axes: intensities in instrument units versus wavelengths in \AA,
   with equivalent Dopplershift from line center of the strongest line
   along the panel tops.
   The second \SiIV\ line (third panel) is drawn at \rev{doubled}
   intensity scale to \rev{offset the factor of two between} their
   \revpar transition probabilities; equal apparent profile heights
   indicate optically thin line formation.
   }
\end{figure*}

EB-1 appeared to be fairly upright, so that the red and blue pixels in
the top row of Fig.~\ref{fig:cutouts_EB1} likely sampled its lower and
upper parts (the spectra \rev{in Fig.~\ref{fig:spectra_EB1}} confirm
this distinction).    
The violet pixel in the second row \rev{of Fig.~\ref{fig:cutouts_EB1}}
sampled the same location on the Sun as the red pixel in the first
row, but during the 2.5 minutes between these samplings EB-1 moved to
the left in its successive flaring, so that the violet pixel sampled
its middle part.
The morphology of the emission patches in Fig.~\ref{fig:cutouts_EB1}
suggests that the 1700\,\AA\ feature came mostly from the lower part
and that the ultraviolet images sampled increasingly more of the upper
part for higher formation temperature (left to right). 
The \Halpha-wing brightness maps only the lower part in the top row,
but in the second row EB-1 appears as an \Halpha-wing brightness patch
that resembles the ultraviolet patches, whereas the 1700\,\AA\ feature
remained dominated by the lower part.

The row-to-row evolution in Fig.~\ref{fig:cutouts_EB1} suggests that
EB-1 grew from the first sample time to the second, while migrating
leftward, and then diminished.
This is confirmed by Fig.~\ref{fig:lightcurves_EB1} which shows the
time behavior of the integrated emission of EB-1 in the \acp{SST},
\acp{IRIS}, and \acp{SDO}/\acp{AIA} imaging.
The integration area was defined appreciably wider than the \Halpha\
detection contour to admit the larger extent of the brightenings in
the \acp{IRIS} slitjaw and \acp{AIA} images
(Fig.~\ref{fig:cutouts_EB1}).
The first part covers the three sample times of
Fig.~\ref{fig:cutouts_EB1} (dashed vertical lines) and shows a rise, a
high peak around 08:32~UT, and a subsequent decay in all diagnostics.
Subsequently there were yet more brightenings, with interesting
differences between the various curves, but unfortunately without
spectroscopic sampling by \acp{IRIS} because EB-1 had drifted off the
slit scan strip.
The \Halpha-core curve shows a high peak a minute after the third
sampling.
Inspection showed that it was from a bright fibril ending of the type
commonly seen at \Halpha\ line center, with many similar ones in the
neighborhood. 
Nothing like the microflare of FAF-1 discussed below.

Figure~\ref{fig:spectra_EB1} shows \acp{SST} \Halpha\ profiles and
\acp{IRIS} ultraviolet-line profiles of EB-1.
The color coding corresponds to the pixel markers in
Fig.~\ref{fig:cutouts_EB1}.
The black profiles are the spatial average over the area specified by
blue frames in Fig.~\ref{fig:fovs}.
These reference profiles serve to gauge the amount of unusual
brightening in the \acp{EB} profiles.

In the first panel of Fig.~\ref{fig:spectra_EB1} the red \Halpha\
profile from the lower part of EB-1 shows characteristic \acp{EB}
signature: excess wings but nothing special in the core.
The blue profile from the upper part shows no \Halpha\ brightening
yet, as already noted in the first row of Fig.~\ref{fig:cutouts_EB1},
but the subsequent violet profile shows considerable \Halpha\ wing
brightening.
This profile also indicates significant core redshift, but this we
attribute to the overlying fibril seen in the line-center movie.
EB-1 contributed only the wing parts outside the steep core flanks. 

If overlying fibrils are opaque in the \Halpha\ core, they must be
much more opaque in \MgIIhk\ for the following reasons.
In \PaperII\ we found already that the core of \CaII~8542\,\AA\ is
affected by overlying fibrils similarly to \Halpha, be it with larger
sensitivity to non-thermal Dopplershifts.
Fibrils that appear opaque in both \Halpha\ and \CaII~8542\,\AA\ must
necessarily be yet more opaque at the centers of the \CaIIHK\ lines,
since these are from the \CaII\ ground state while the 8542\,\AA\ line
is from an excited level.
Such fibrils must then be 18 times (Mg/Ca abundance ratio) more opaque
yet at the centers of the \MgIIhk\ lines.

The \MgIIhk\ cores in the bottom right panel of
Fig.~\ref{fig:spectra_EB1} indeed show only a small response and the
2796\,\AA\ slitjaw images in Fig.~\ref{fig:cutouts_EB1} show less
\acp{EB} brightening than the other diagnostics.
However, overlying fibrils must also become transparent further out in
the \hk\ wings, just as in the \Halpha\ wings.
The violet \hk\ profiles in Fig.~\ref{fig:spectra_EB1} indeed show
outer-wing brightening.
The 2796\,\AA\ light curve in Fig.~\ref{fig:lightcurves_EB1} shows a
peak around 08:32~UT from these broader \MgIIk\ wings.

In contrast to the fibril-dominated \Halpha\ and \MgIIhk\ cores, there
is large response to EB-1 in the \SiIV\ and \CII\ profiles in
Fig.~\ref{fig:spectra_EB1} and also in the \MgII\ triplet lines
between \hk.
They all show a clear progression of excess emission for the red,
blue, and violet samplings, again suggesting that the upper part of EB-1
was hotter than the lower part and became hotter with time.
From red to blue the \CII\ lines also became much wider.
Only their central self-absorption dips remained unaffected.
These are probably also fibrilar.

In addition to this brightening, the profiles of all these lines show
marked asymmetries with very good correspondence between them.
The redshifts of the red profiles from the lower part suggest
downflow, the blueshifts of the blue and violet profiles from the upper
part suggest upflow.  
These patterns provide direct evidence for the presence of a
bi-directional flow, as discussed earlier for \acp{EB}s by
\citetads{2007Sci...318.1591S}, 
\citetads{2008PASJ...60...95M}, 
\citetads{2009A&A...508.1469A} 
and in \PaperI.

The blue \SiIV\ peaks show blueshifts of roughly 30~\kms.
More precise fits with double Gaussians to reproduce their asymmetry
gave blueshift magnitudes of about 45~\kms\ for the main (i.e.,
highest intensity) components.
The violet \SiIV\ peaks for the later sampling show smaller blueshifts
(about 15~\kms) but this was the middle part of the \acp{EB}, not its
top, due to its Eastward progression (Fig.~\ref{fig:cutouts_EB1}).

Thus, there is no point in inspecting (or modeling) the cores of
\Halpha\ and \MgIIhk\ to study \acp{EB} behavior, but the striking
agreement in Doppler asymmetries for the \MgII\ triplet, \SiIV, and
\CII\ lines suggests that these sample the underlying \acp{EB} without
fibrilar obscuration.
These \acp{IRIS} lines thus provide diagnostics in which \acp{EB}s are
``unveiled''. 

\begin{figure}
\centerline{\includegraphics[width=1.0\columnwidth]{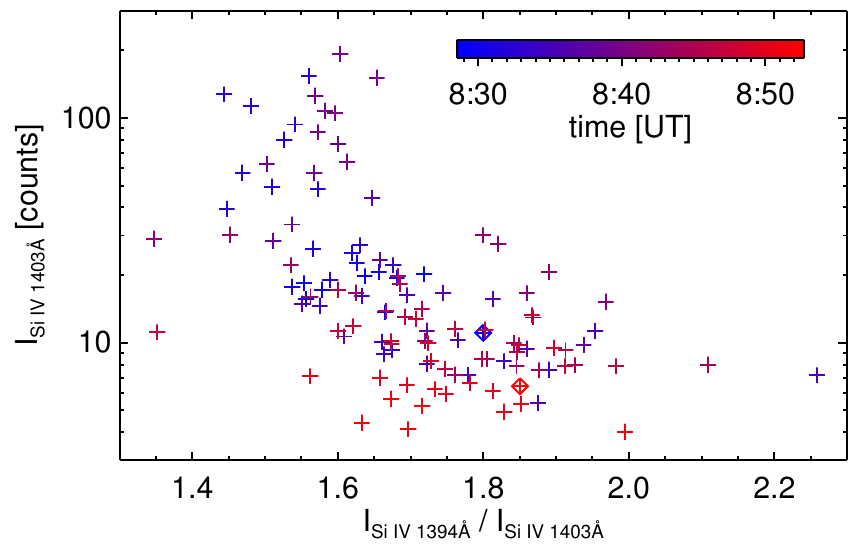}}
   \caption[]{\label{fig:Si_ratio_EB1} 
   \rev{Peak} intensity of the \SiIV~1403\,\AA\ \rev{profile} against
   its ratio with the \rev{peak} intensity of the \SiIV~1394\,\AA\
   \rev{profile for} all spectral samplings of EB-1.
   \rev{The color coding specifies observation time.
   The two diamonds specify the start and end samplings.}
   }
\end{figure}

In addition, these \acp{EB}-unveiling diagnostics differ clearly in
how they sample EB-1.
The \SiIV\ lines indicate a \rev{roughly} optically thin feature for
the red and blue profiles even at their centers, because each pair
reaches \rev{similar height} in the second and third panels of
Fig.~\ref{fig:spectra_EB1} and \revpar shows no flattening or dip
\rev{in the line cores}.
Since the plot scales of the \SiIV\ panels differ by the transition
probability ratio, such apparent height equality suggests optically
thin formation.
For the violet profiles this is not the case, suggesting that the
increased emissivity went together with larger EB-1 opacity.

This thickness measure is quantified in Fig.~\ref{fig:Si_ratio_EB1}
which plots the intensity of one \SiIV\ peak against the peak ratio.
They were measured by averaging the profiles over all spatial-temporal
samplings of pixels within the bright EB-1 patch in the 1400\,\AA\
slitjaw images that correspond to each successive \acp{IRIS} scan, and
\rev{smoothing the top of each averaged profile to measure} its
maximum intensity.

For an \revpar 
optically thin cloud \rev{without background irradiation} the emergent
doublet intensities equal the local emissivities times the geometrical
thickness and obey the transition probability ratio of 2.
For thicker features the ratio reduces, reaching unity for an opaque
cloud with constant source function and then yielding flat-topped
profiles (Fig.~2.2 of
\citeads{2003rtsa.book.....R}). 
Figure~\ref{fig:Si_ratio_EB1} suggests that EB-1 was mostly
\rev{neither thin nor thick, but ``thinnish'' with ratio values
between 1.9 and 1.5 that typically correspond to thicknesses 0.3 and
1.6, respectively.
The sample-time coding (color) suggests that EB-1 started and ended
optically thin, but was generally somewhat thicker at high peak
intensities in between.} 

When such features 
become much thicker than the photon mean free path, internal resonance
scattering tends to cause outward source function decline and a
corresponding central profile dip that is commonly called
self-absorption.  
Such dips are seen in the \CII\ lines and also in the \MgII\ triplet
lines.
However, their \revpar peak \rev{heights and} asymmetries correspond
very well to the \SiIV\ profiles. \revpar \rev{Their peaks rise in
concert and the higher peak of each profile is on the side to which
the \SiIV\ peaks are shifted.
This good correspondence suggests} that also these peaks sampled EB-1
without obscuration from overlying fibrils.
The latter caused only the central dips because these show no
Dopplershifts.
Since the outer wings still show about similar intensities as function
of wavelength separation from the line centers, the peaks and wings
seem to sample the \acp{EB} in optically thick fashion (otherwise they
would also differ a factor of two, the transition probability ratio).  
For an optically thick feature the profiles represent
Eddington-Barbier mapping of the source function at monochromatic
optical depth $\tau_\lambda\is1$.
In this case the different Dopplershifts of the upper and lower parts
of EB-1 affect the optical depth scaling and produce the peak
asymmetries.
Thus, these unveiled \acp{IRIS} diagnostics provide both thin and
thick \acp{EB} sampling.

The final features of interest in Fig.~\ref{fig:spectra_EB1} are the
line blends.
The \SiIV\ and \CII\ lines are too narrow to reach the nearby \FeII\
and \NiII\ lines (rest wavelengths indicated by dashed vertical
lines), but the blue wing of \MgIIk\ and the raised overlapping wings
between \hk\ contain strong \MnI\ absorption blends at 2795.64,
2799.09, and 2801.91\,\AA\ in the red lower-part sampling (best seen
per zoom-in with a pdf viewer).
They are weak or absent in the other samplings.
We attribute them to foreground upper-photosphere gas crossed by the
slanted line of sight towards EB-1 that is not part of the phenomenon
and indeed imparts no obvious Dopplershift.
A line of sight to the lower part then passes through the \MnI\
formation layer, a line of sight towards the upper part catches less
or none.

The \MnII\ blend at 2794.72\,\AA\ shows up with interesting profiles:
as a self-reversed line in the red lower-part sampling, absent in the
blue upper-part sampling, but appearing with a blue-peaked profile
similar to the \MgII\ triplet lines at 2798.82\,\AA\ in the later violet
upper-part sampling. 
This similarity indicates sampling of the \acp{EB} itself.

In summary, the \acp{IRIS} diagnostics provide an informative,
understandable, and self-consistent view of EB-1 that fits very well
with our earlier \acp{EB} descriptions.  
EB-1 appeared as a photospheric below-the-fibrilar-canopy heating
event with upward progression with time, larger heating higher up, and
unmistakable bi-directional jet signature.

Are these \acp{EB} spectra similar to the \acp{IB} spectra of
\citetads{2014Sci...346C.315P}? 
While the \CII\ and \SiIV\ lines do show brightening, also with
bi-directional Dopplershift signatures, their wings are not
extravagantly wide.
This may be a matter of timing.
It is a pity that \acp{IRIS} did not sample its aftermath, as is
demonstrated by the next example.

\subsection{Details for EB-2}

\begin{figure}
  \centerline{\includegraphics[width=\columnwidth]{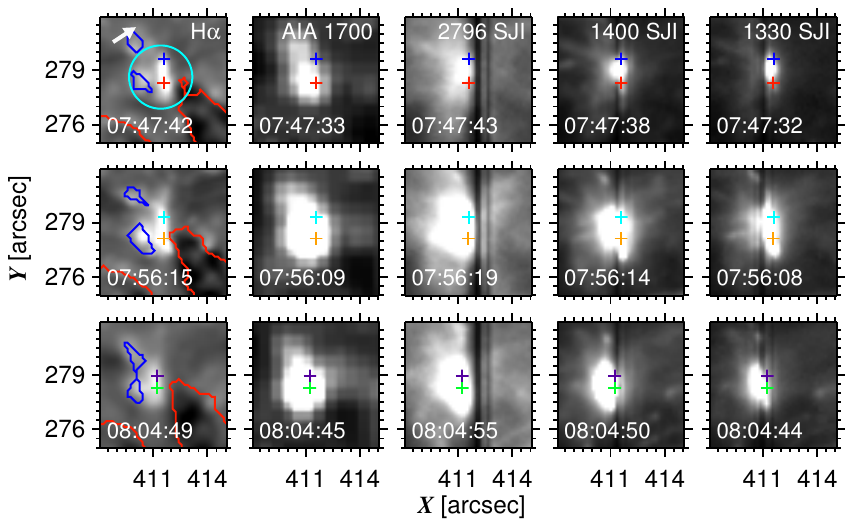}}
    \caption[]{\label{fig:cutouts_EB2} 
    Time evolution of EB-2 in the format of
    Fig.~\ref{fig:cutouts_EB1}\rev{, except that
    the red and blue contours in the first column are based on
    CRISP \FeI~6302\,\AA\ Stokes $V$ data (thresholded at
    $\pm$450\,counts).}
    In each panel the pair of pixel markers specifies the sample
    locations of simultaneously recorded spectra shown in
    Fig.~\ref{fig:spectra_EB2}.
    The temporal separation between rows is about 8~min, larger than
    in Fig.~\ref{fig:cutouts_EB1}.
    It corresponds to successive \acp{IRIS} rasters.}
\end{figure}

\begin{figure}
\centerline{\includegraphics[width=\columnwidth]{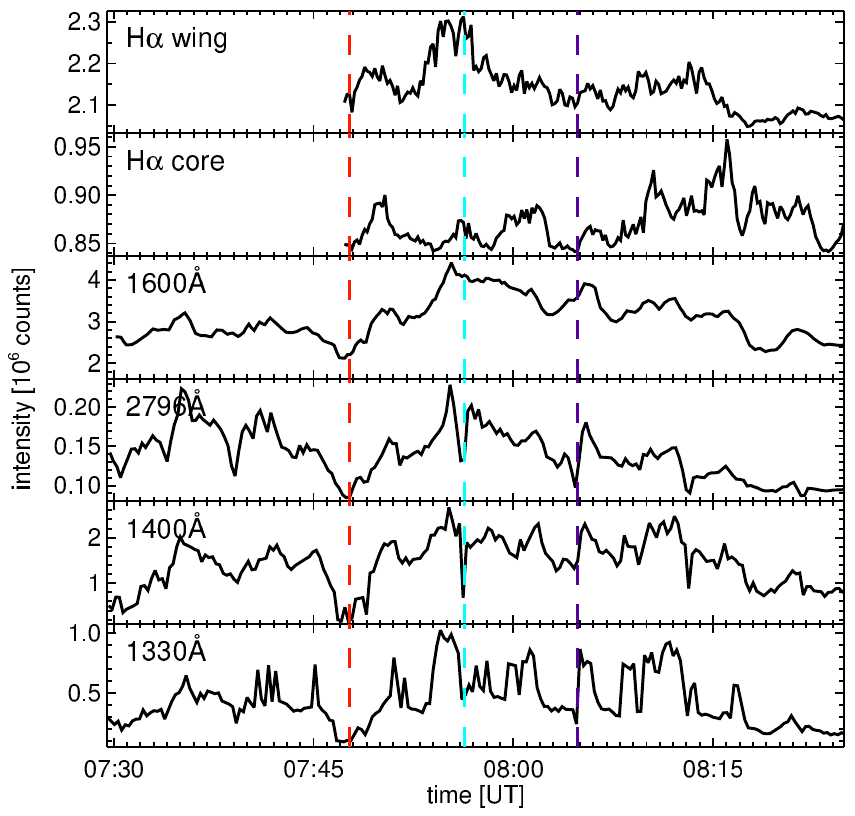}}
  \caption[]{\label{fig:lightcurves_EB2} 
  Light curves for EB-2 in the format of
  Fig.~\ref{fig:lightcurves_EB1}.
  In this case the integration aperture had a diameter of 3.5~arcsec
  (1.5 times the maximum diameter of the \Halpha\ \acp{EB} detection
  contour)\rev{, as indicated in the first panel of
  Fig.~\ref{fig:cutouts_EB2}}.
  The three dashed vertical lines correspond to the sampling times of
  the three rows in Fig.~\ref{fig:cutouts_EB2}.  
  There are corresponding dips in the lower three curves from the slit
  presence over the feature.
  }
\end{figure}

\begin{figure*}
 \centerline{\includegraphics[width=\textwidth]{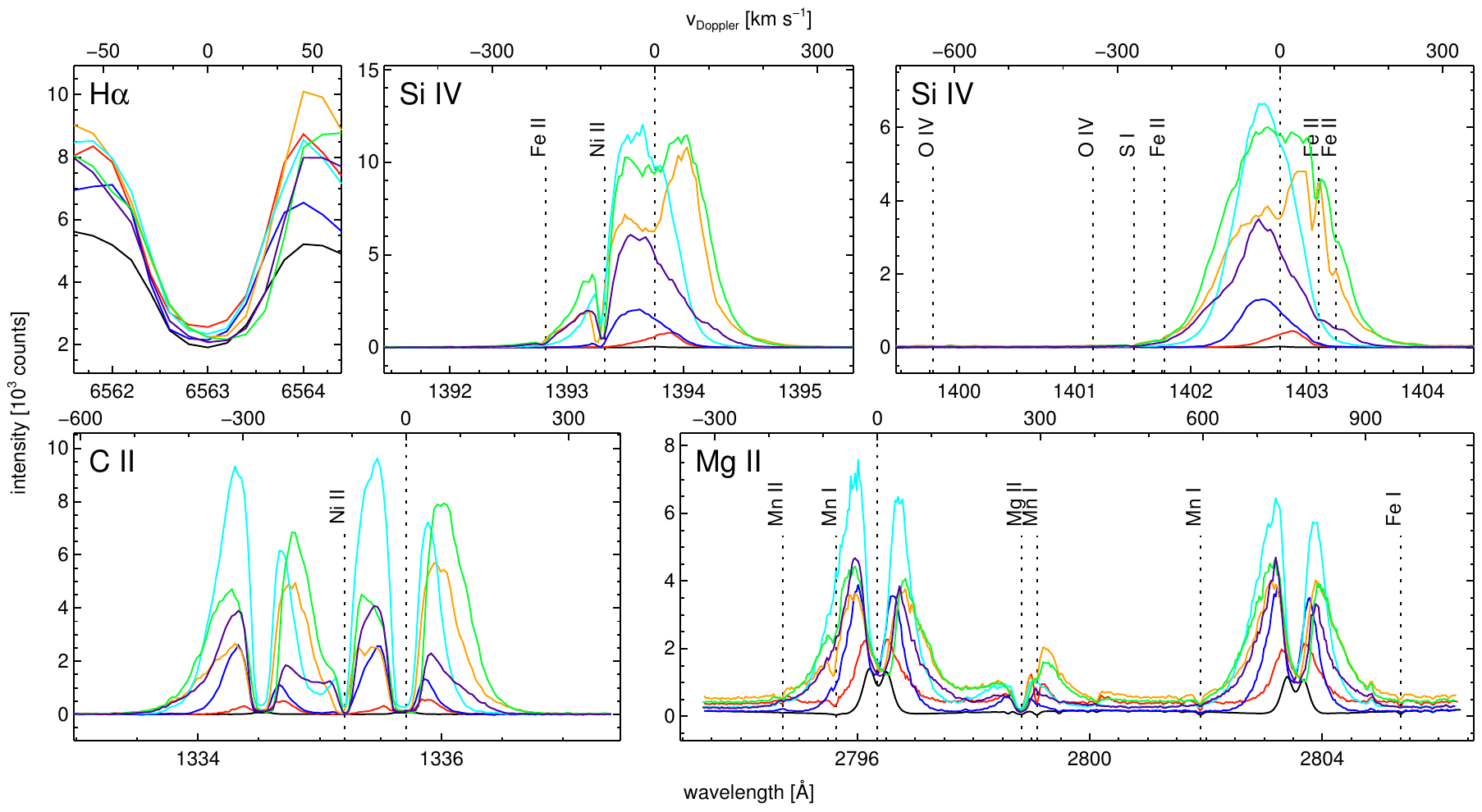}}
   \caption[]{\label{fig:spectra_EB2} 
   \acp{CRISP} and \acp{IRIS} spectra of EB-2 at the marker positions in
   Fig.~\ref{fig:cutouts_EB2}, with the same color coding.
   The red and blue profiles were taken at 07:47:38~UT, the orange and
   cyan profiles at 07:56:14~UT, the green and violet profiles at
   08:04:45~UT.
   Format as for Fig.~\ref{fig:spectra_EB1}, except that the \Halpha\
   panel is compressed in wavelength to make place for wider \SiIV\
   1394\,\AA\ profiles conserving their Doppler velocity scale.}
\end{figure*}

EB-2 was sampled in \acp{IRIS} spectroscopy during a longer period,
but only intermittently due to the slow raster repeat at 8.6-minute
cadence.
Figures~\ref{fig:cutouts_EB2}--\ref{fig:spectra_EB2} display results
for EB-2 in the same format as
Figs.~\ref{fig:cutouts_EB1}--\ref{fig:spectra_EB1}. 
This \acp{EB} had the advantage that it appeared aligned along the
\acp{IRIS} slit, so that its top and bottom were spectroscopically
sampled at the same time in each row of Fig.~\ref{fig:cutouts_EB2}.
The alignment was fortuitous since EB-2 was tilted considerably away
from the local vertical in its azimuthal orientation (angle with the
arrow in the first panel of Fig.~\ref{fig:cutouts_EB2}).

Inspection of the \acp{SST} \Halpha\ sequences showed an unmistakable
large, repetitive \acp{EB} flame.
\revpar
EB-2 was already present at the start of the \acp{SST} observations at
07:47~UT, quickly brightened, and remained nearly continuously present
in the \acp{AIA} image and \acp{IRIS} slitjaw sequences until
08:29~UT.
These image sequences and the corresponding light curves in
Fig.~\ref{fig:lightcurves_EB2} suggest that there was also preceding
\acp{EB} activity during fifteen minutes before the \acp{SST} start.
A longer-duration \acp{AIA} 1700\,\AA\ sequence suggests strong
repetitive \acp{EB} activity at the same location already from
06:36~UT onward.   
The \acp{HMI} magnetogram sequence shows very fast streaming with an
extended patch of white polarity running into a fairly large patch of
black polarity that vanished gradually and was gone by 08:30~UT.
The \acp{SST} magnetograms have higher spatial resolution but only at
the best-seeing moments; the homogeneity of the \acp{HMI} sequence
makes it more suited to follow such pattern changes.

The hotter \acp{AIA} diagnostics (304, 171, 193\,\AA) show nothing
special happening at the site of EB-2 during the whole period.
EB-2 thus seems a bonafide \acp{EB}, but repeating over unusual
duration thanks to continued supply of opposite-polarity fields.

The top row of Fig.~\ref{fig:cutouts_EB2} with red and blue pixel
markers shows EB-2 at the start of the \acp{SST} observations (first
vertical line in Fig.~\ref{fig:lightcurves_EB2}).
The corresponding red and blue \SiIV\ and \CII\ profiles in
Fig.~\ref{fig:spectra_EB2} appear similar in shape to those in
Fig.~\ref{fig:spectra_EB1}, but note the differences in intensity
scales: the blue and red \SiIV\ profiles reach nearly \rev{20} times higher
count values than for EB-1, more than expected from the doubled
exposure time.
They are also much wider.
The red \CII\ profiles are still well separated, but the blue profiles
nearly overlap.
This is not seen in quiet-Sun spectra
(\citeads{1978ApJ...222..333L}). 

The red and blue profiles nevertheless still display similar profiles
as EB-1: clean humps with the blue ones from the upper part reaching
higher intensities than the red ones from the lower part, and with
redshifts for the lower sampling and blueshifts for the upper sampling
that again correspond very well between the \SiIV, \CII, and \MgII\
triplet lines.
As in EB-1, the 1700\,\AA\ brightness patch in the top row of
Fig.~\ref{fig:cutouts_EB2} favors the bottom part of EB-2 while the
ultraviolet images indicate higher formation for hotter diagnostics.
These EB-2 results are in excellent agreement with our EB-1 findings.

However, the red and blue profiles sampled only the start-up of EB-2.
It became much larger and brighter afterwards
(cf.~Figs.~\ref{fig:cutouts_EB2} and \ref{fig:lightcurves_EB2}).
Figure~\ref{fig:spectra_EB2} adds lower-and-upper spectral sample
pairs also at the times when the \acp{IRIS} slit passed EB-2 again
around 07:56~UT and 08:05~UT, respectively (second and third rows of
Fig.~\ref{fig:cutouts_EB2}, with orange and cyan lower/upper markers
in the second row and green and violet lower/upper markers in the
third row).
The \SiIV\ and \CII\ lines grew considerably in intensity and
developed more complex profiles that do show \acp{IB} signatures: wide
wings, cores with complex structure, and deep blends.

The orange and green lower-part samplings show the most complex
\rev{double-peaked} \SiIV\ profiles in
Fig.~\ref{fig:spectra_EB2}. \revpar \rev{These profiles come closest 
in shape to the double-peaked profiles in 
\citetads{2014Sci...346C.315P}, 
which were interpreted as signature of a bi-directional jet.
Here, t}he small \SiIV\ line-center dips may represent self-absorption
in stationary gas, but also just lack of stationary gas in
optically-thin emissivity mapping of two counter streams as blue- and
redshifted profile humps.
\rev{In contrast to the similarity of the two \SiIV\ 
profiles in \citetads{2014Sci...346C.315P}, 
the green and orange \SiIV\ profiles of EB-2 differ between the two lines.}

The Dopplershift patterns are again consistent between the different
unveiled diagnostics.
All lower-part samples (red, orange, green) show redshifts domination,
the upper-part samples (blue, cyan, violet) single blueshifted peaks,
although with redward profile tails. 
Most profiles show ragged tops.
Such core raggedness is further discussed in
Sect.~\ref{sec:discussion}.

The \NiII\ blends at 1393.32\,\AA\ in the stronger \SiIV\ line and at
1335.20\,\AA\ between the \CII\ lines are very pronounced in all but
the red samplings.
The \FeII\ blends at 1403.10\,\AA\ and 1403.26\,\AA\ in the red wing
of the weaker \SiIV\ line are present in the orange and green
lower-part samplings, and weakly in the final violet upper-part
sampling.
These blends generally show blueshifts, larger in the orange than in
the subsequent green sampling.
They suggest upward, decelerating motion of cool gas along the line of
sight to the \acp{EB}.

The three lower-part samples (red, orange, green) show the \MnI\
blends in the \MgIIhk\ wings, strongest in the red start-up lower
sampling, whereas they are not present in the three upper samplings. 
We again attribute these blends to undisturbed upper-photosphere gas
along slanted lines of sight to the \acp{EB} foot, with lines of sight
towards the upper part passing over the \MnI\ formation layer.
In this case the \MnII\ line is only weakly present in the lower-part
samplings, without \acp{EB} sampling.

The core of \Halpha\ remained similar in the various samplings, again
indicating domination by overlying fibrils.  
The same holds for the line-center dips of \MgIIhk.
The green \Halpha\ and \hk\ cores show similar redshift.

The \MgIIhk\ peaks brightened considerably in the cyan upper-part
sampling (the dip in the 2796\,\AA\ light curve in
Fig.~\ref{fig:lightcurves_EB2} is due to the slit).
This curve tracks the 1700\,\AA\ light curve fairly well.
Since also the peak asymmetries correspond with those of the \CII\
lines the fibrilar obscuration may have been thinner for this
sampling.
The cyan \Halpha\ core is also relatively narrow.

The outer wings of the \MgIIhk\ seem to sample EB-2 in optically-thick
manner because they have about equal intensities in the two lines
(which also have a transition probability ratio of two). 

We conclude that EB-2 showed EB-1-like profiles in its onset and later
developed more outspoken \acp{IB} signatures.
Whether EB-1 did the same we do not know, but EB-2 shows that strong
\acp{EB} activity can indeed produce \acp{IB}-type spectra as suggested by
\citetads{2014Sci...346C.315P}. 

The similarity of the orange and green lower-part profiles and the
similarity of the cyan and violet upper-part profiles at 8.6~min
sampling delay suggests that the feature persisted over long duration.

Such a hot signature was not seen at the onset of EB-2, implying that
the preceding hour of \acp{EB} activity suggested by the \acp{AIA}
1700\,\AA\ movie had not left one by that time.

\subsection{Details for EB-a, EB-b, EB-c}

\begin{figure}
  \centerline{\includegraphics[width=\columnwidth]{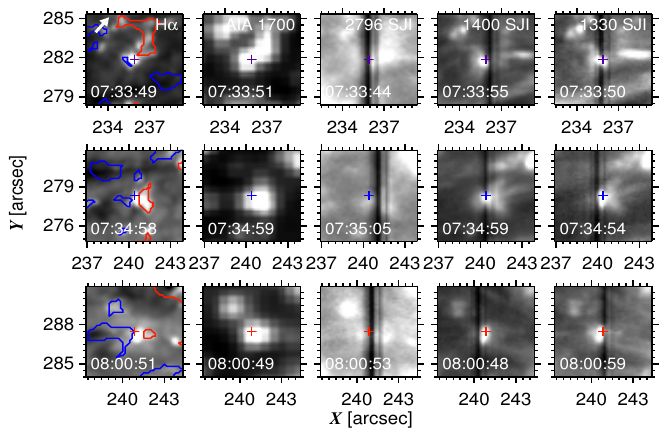}}
    \caption[]{\label{fig:cutouts_EBabc} 
    Image cutouts for EB-a ({\it top row\/}), EB-b ({\it middle
    row\/}) and EB-c ({\it bottom row\/}).
    The format is the same as for Fig.~\ref{fig:cutouts_EB2}, except
    that the rows are for different locations and all panels were
    therefore bytescaled individually.
    \rev{The red and blue polarity contours represent thresholds of +100
    and -250\,counts, respectively.}
    }
\end{figure}

\begin{figure*}
 \centerline{\includegraphics[width=\textwidth]{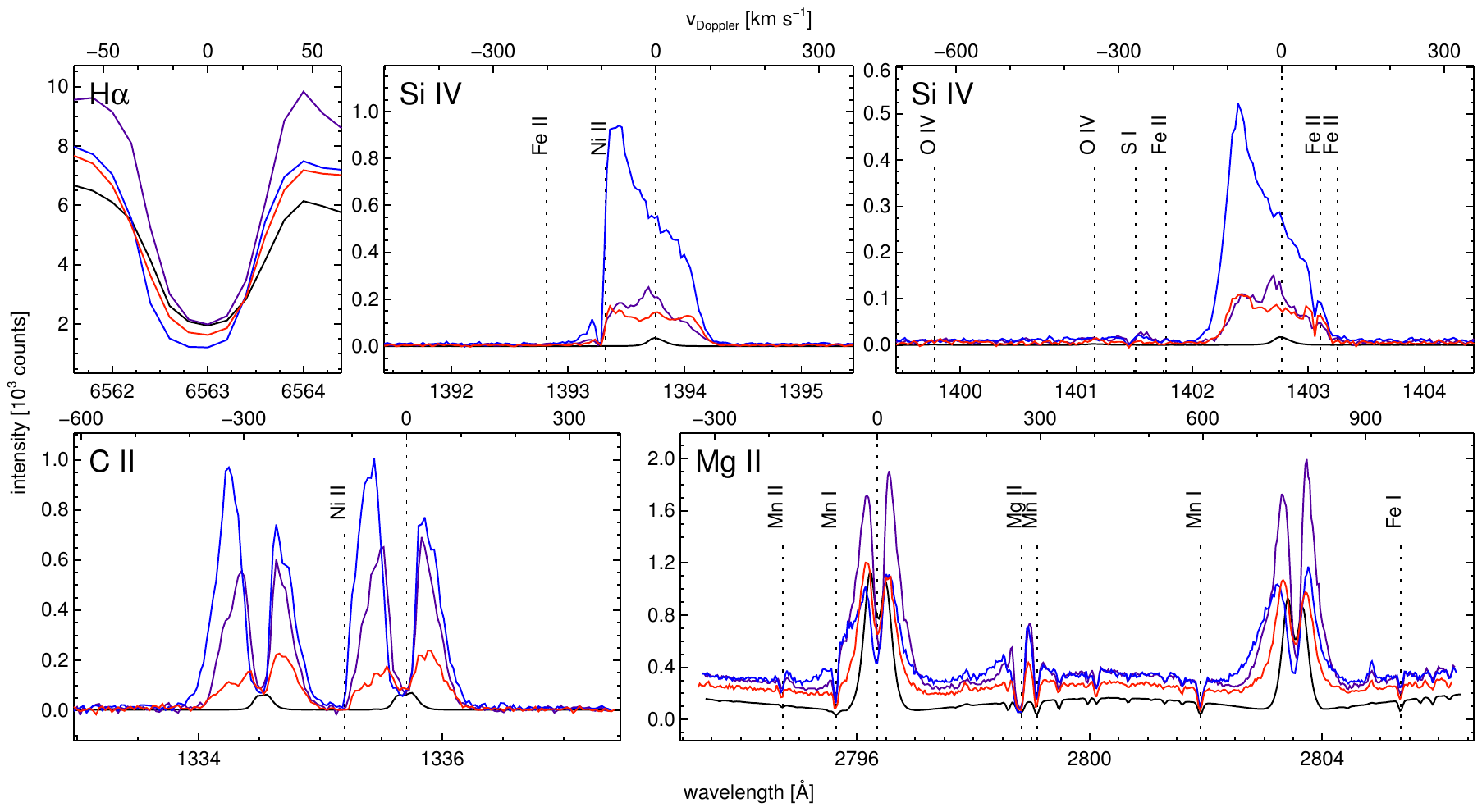}}
   \caption[]{\label{fig:spectra_EBabc} 
   \acp{CRISP} and \acp{IRIS} spectra of EB-a ({\it violet\/}), EB-b
   ({\it blue\/}), and EB-c ({\it red\/}) in dataset~2 at the marker
   positions in Fig.~\ref{fig:cutouts_EBabc} and with the same color
   coding.
   Format as for Fig.~\ref{fig:spectra_EB1}.
   }
\end{figure*}

EB-1 and EB-2 were well-defined strong \acp{EB}s.
We now turn to the weaker or questionable EB-a, EB-b, EB-c in
dataset~2.
Figures~\ref{fig:cutouts_EBabc} and \ref{fig:spectra_EBabc} show their
sampling and spectral profiles in the format of
Figs.~\ref{fig:cutouts_EB1} and \ref{fig:spectra_EB1}.

The \acp{HMI} magnetogram sequence displays a complex region with much
streaming motion in which opposite-polarity patches canceled
frequently.
There were strong \Halpha\ \acp{EB}s that appeared as obvious
1700\,\AA\ ones, but these were unfortunately not sampled by
\acp{IRIS}.
As noted above, we lowered the \Halpha\ discrimination level to
include weaker events that were sampled by \acp{IRIS} because their
\Halpha\ morphology indicated \acp{EB} nature rather than \acp{MC}
nature, although such recognition was hampered by more vertical
viewing than in datasets 1 and 3.
In the \acp{AIA} 1700\,\AA\ movie none of these appeared as an obvious
\acp{EB} (even though EB-b passed the \PaperII\ 1700\,\AA\ criteria).
For example, in the first two panels of Fig.~\ref{fig:cutouts_EBabc}
the \Halpha\ feature appears EB-like, but the neighboring normal
\acp{MC}s appear as bright in 1700\,\AA.

The three \acp{EB}s we selected for display are the ones with the highest
intensities in the \SiIV\ lines.
They are sufficiently high to confirm that these candidates were not
pseudo-EBs (ordinary \acp{MC}s; if they were, \acp{IRIS} would show
such profiles from network everywhere). 
We therefore present these as non-suspect but weak \acp{EB}s.

In Fig.~\ref{fig:spectra_EBabc} the \SiIV\ lines reach \rev{about twice as high} 
values as EB-1 in Fig.~\ref{fig:spectra_EB1} \rev{which could be explained from
the doubling of the exposure time)}, but these \SiIV\ profiles are more complex.
The apparent \SiIV\ height equality between the differently scaled
panels again indicates \rev{thinnish formation, with the} non-Gaussian
shapes suggesting viewing through multiple Dopplershifted components.
This strengthens our judgment from \PaperI\ and \PaperII\ that slanted
\acp{EB} viewing helps to diagnose \acp{EB} formation.
In Figs.~\ref{fig:spectra_EB1} and Fig.~\ref{fig:spectra_EB2} it did
by spatially separating the different Dopplershift signatures of the
lower and upper parts of the bi-directional jet.  
Here, these likely mixed together along the line of sight in all three
\acp{EB}s.

Thus, the \SiIV\ profiles in Fig.~\ref{fig:spectra_EBabc} suggest that
the line of sight sampled both a blueshifted upper part and a
redshifted lower part, with the upper part again hotter for EB-b (blue
profiles) but with about equal contributions for the other two.
All \SiIV\ cores show small-scale raggedness
(Sect.~\ref{sec:discussion}).

The \CII\ profiles are similar to those of EB-1 in
Fig.~\ref{fig:spectra_EB1} but with closer peak equalities that again
suggest bi-modal sampling.
For EB-b the blue profile of the \MgII\ triplet lines shows opposite
asymmetry to the blue \SiIV\ and \CII\ profiles, suggesting peak
formation in the lower part similar to the three lower-part samples of
EB-2 in this line in Fig.~\ref{fig:spectra_EB2}.

The various blends are markedly present in all three samplings.
The \FeII\ and \NiII\ blends again show substantial blueshifts, the
\MnI\ and \MnII\ lines in the last panel none. 
In top-down viewing as suggested by the jet mixing, the latter must be
from cool upper-photosphere-like gas above the \acp{EB}s, suggesting
that the \acp{EB} flames did not reach high into the atmosphere.

The upshot is that these three weak \acp{EB}s adhere to the pattern
set by EB-1 and EB-2 in their start-up phases, but without spatially
resolving the bi-directional jets.
It is better to observe \acp{EB}s away from disk center, which also
increases \acp{EB} contrasts over \acp{MC}s at 1700\,\AA.

\subsection{Details for FAF-1}

\begin{figure}[t]
  \centerline{\includegraphics[width=\columnwidth]{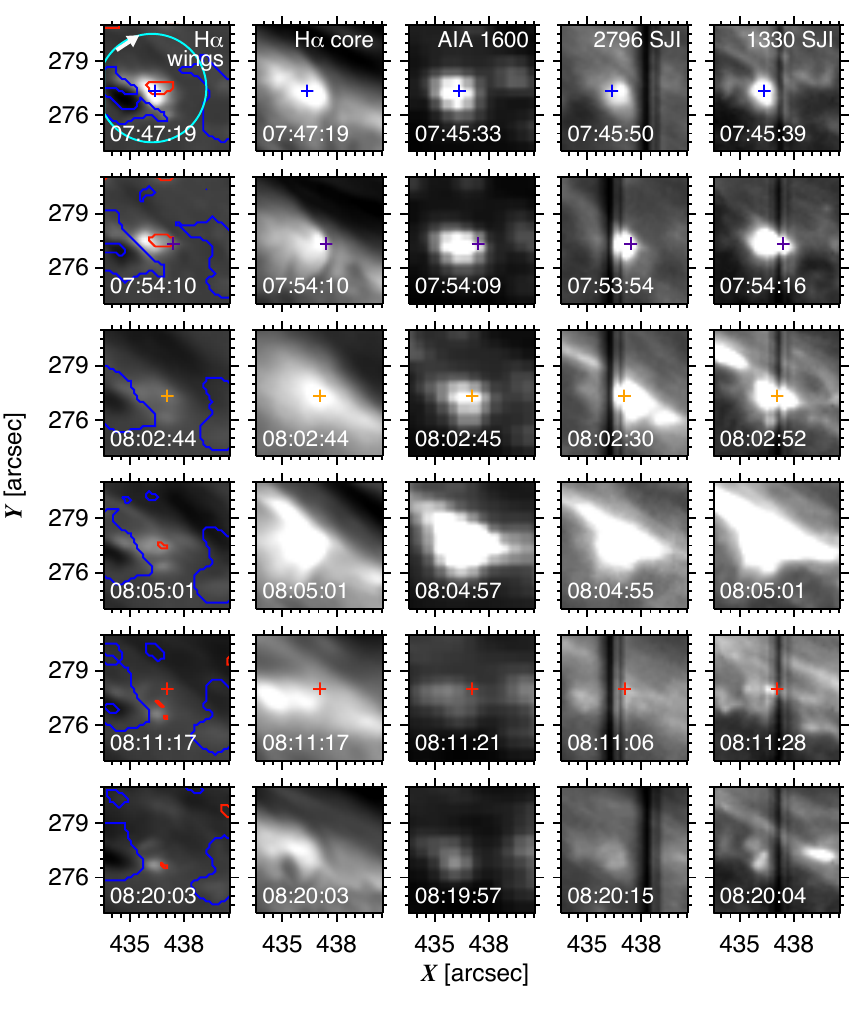}}
    \caption[]{\label{fig:cutouts_FAF1} 
    Image cutouts for FAF-1.
    Format as for Fig.~\ref{fig:cutouts_EB2}, except that \acp{SST}
    \Halpha\ core images are added in the second column, \acp{AIA}
    1600\,\AA\ images instead of 1700\,\AA\ images are shown in the
    third column, and no 1400\,\AA\ slitjaw images are shown because
    they were all nearly identical to the 1330\,\AA\ ones.
    \rev{The red and blue polarity contours represent thresholds of +100
    and -450\,counts, respectively.}
    The sample times correspond to five successive (8.6~min apart)
    slit samplings of this location in the \acp{IRIS} raster pattern,
    plus the moment of the \Halpha\ microflare (4th row). 
    The bytescaling is the same along columns.}
\end{figure}

\begin{figure}[t]
\centerline{\includegraphics[width=\columnwidth]{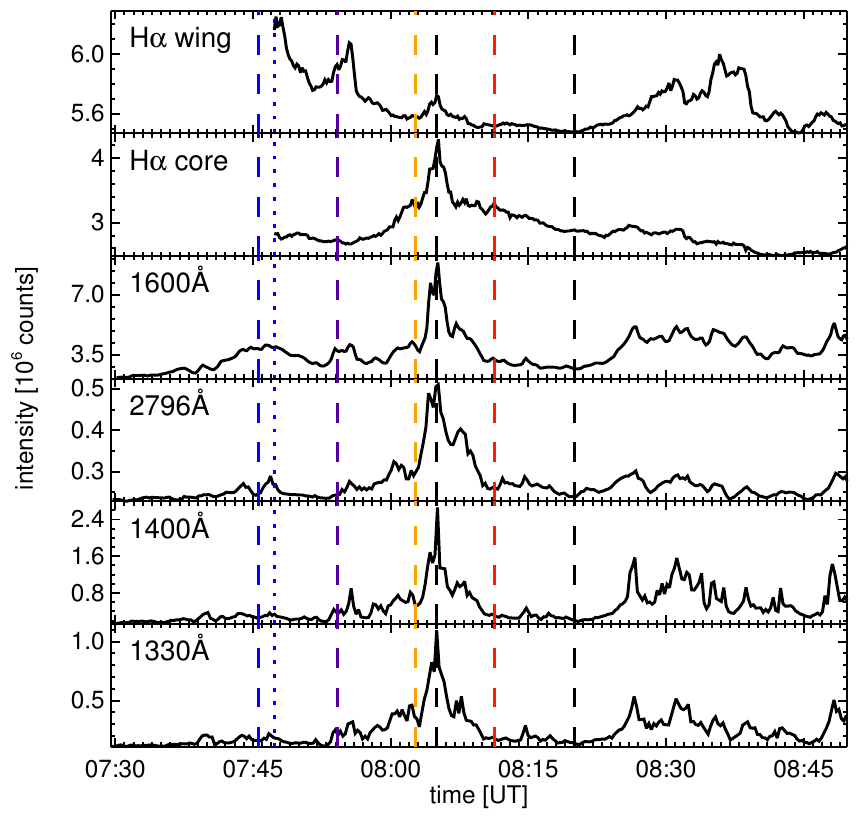}}
  \caption[]{\label{fig:lightcurves_FAF1} 
  Light curves for FAF-1 in the format of
  Fig.~\ref{fig:lightcurves_EB1}.
  The integration aperture had a diameter of 6.0~arcsec\rev{, as
  indicated in the first panel of Fig.~\ref{fig:cutouts_FAF1}}.
  
  The vertical lines correspond to the sampling times of the rows in
  Fig.~\ref{fig:cutouts_FAF1}.
  The dotted one is for the first \Halpha\ sampling, slightly offset
  from the first \acp{IRIS} sampling because the \acp{SST} observation
  started at 07:47~UT.
  }
\end{figure}

\begin{figure*}
 \centerline{\includegraphics[width=\textwidth]{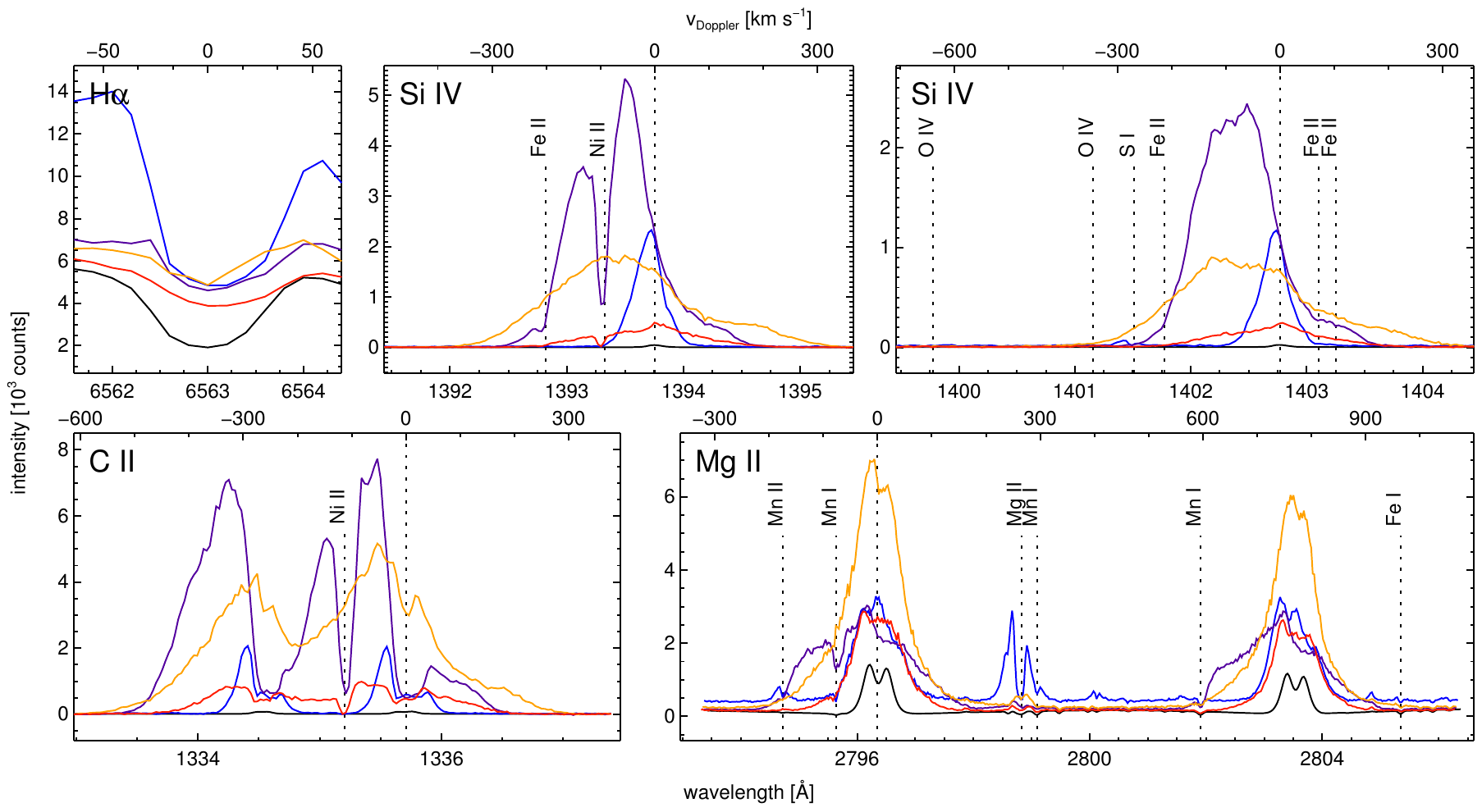}}
   \caption[]{\label{fig:spectra_FAF1} 
   \acp{CRISP} and \acp{IRIS} spectra of FAF-1 at the pixels marked in
   Fig.~\ref{fig:cutouts_FAF1}.
   Format as for Fig.~\ref{fig:spectra_EB1}.
   }
\end{figure*}

\begin{figure}[t]
  \centerline{\includegraphics[width=\columnwidth]{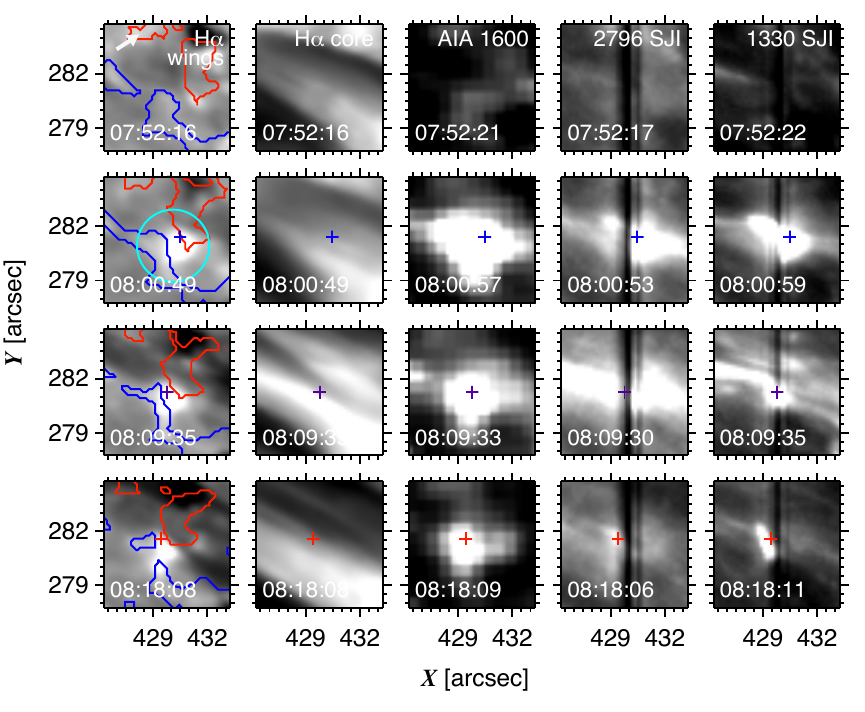}}
    \caption[]{\label{fig:cutouts_FAF2} 
    Image cutouts for FAF-2 in the format of
    Fig.~\ref{fig:cutouts_FAF1}\rev{, with opposite polarity
    contours at thresholds of $\pm$500\,counts}.
    }
\end{figure}

\begin{figure}[t]
\centerline{\includegraphics[width=\columnwidth]{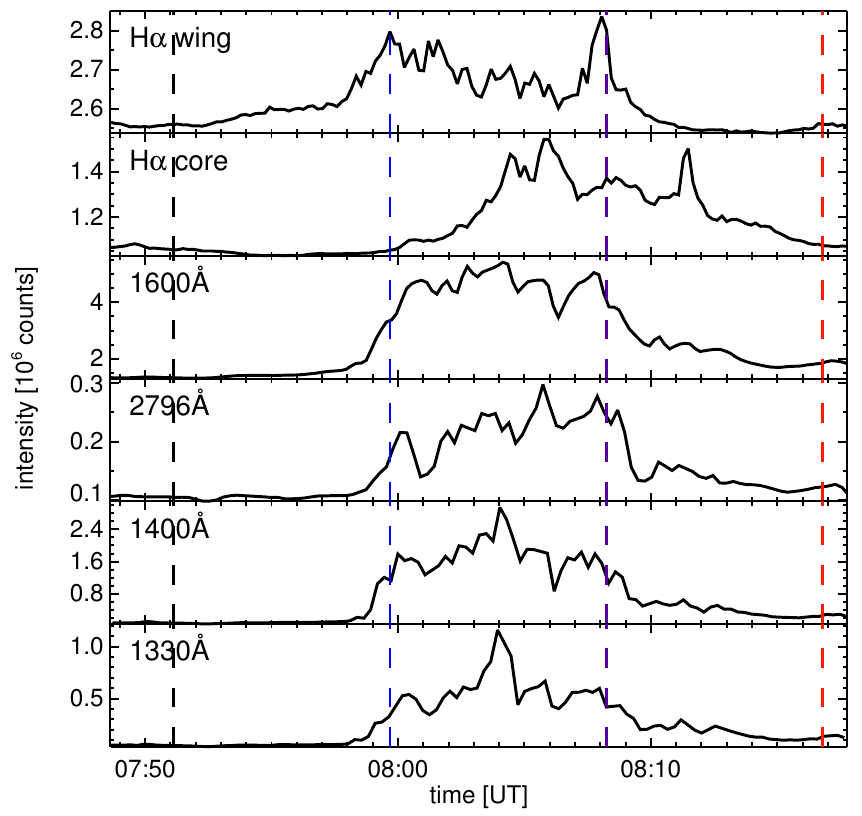}}
  \caption[]{\label{fig:lightcurves_FAF2} 
  Light curves for FAF-2 in the format of 
  Fig.~\ref{fig:lightcurves_FAF1}.
  The integration aperture had a diameter of 4.0~arcsec\rev{, as
  indicated in the second panel of Fig.~\ref{fig:cutouts_FAF2}}.
  
  }
\end{figure}

\begin{figure*}
 \centerline{\includegraphics[width=\textwidth]{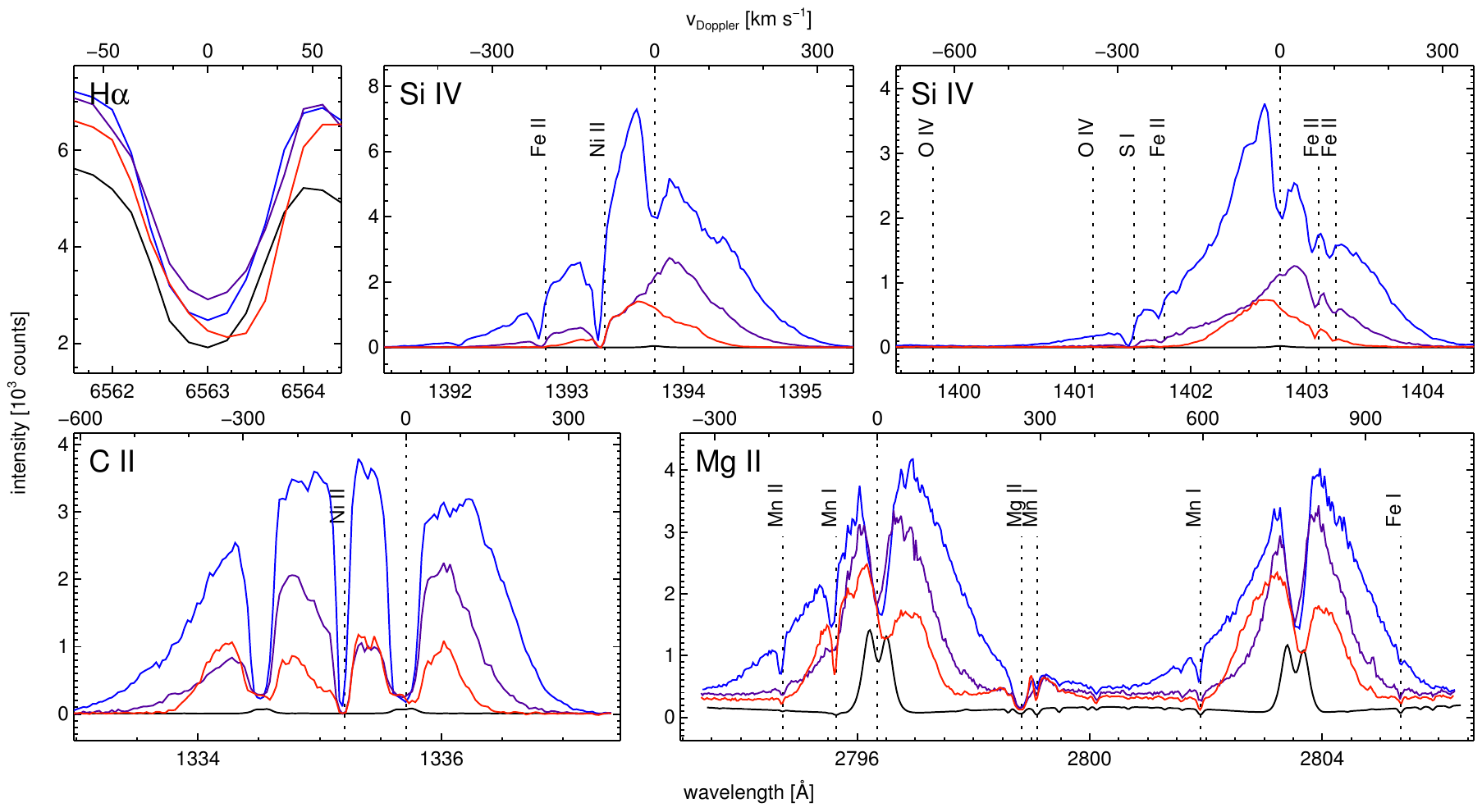}}
   \caption[]{\label{fig:spectra_FAF2} 
   \acp{CRISP} and \acp{IRIS} spectra FAF-2 at the pixels marked in
   Fig.~\ref{fig:cutouts_FAF2}.
   Format as for Fig.~\ref{fig:spectra_EB1}.
   }
\end{figure*}

In dataset 3 a string of very bright \acp{FAF}s appeared in the lower
part of the field of view in Fig.~\ref{fig:fovs}, East of EB-2.
We selected two for display in
Figs.~\ref{fig:cutouts_FAF1}--\ref{fig:spectra_FAF2}.
Their locations are specified in Fig.~\ref{fig:fovs}.

Blinking the \acp{AIA} 1700\,\AA\ and \acp{HMI} magnetogram movies
shows that FAF-1 started when a small patch of black polarity ran fast
from afar to merge with a larger black one adjacent to yet a larger
patch of white polarity; all black polarity then vanished.  
FAF-2 occurred next to a fairly large black-polarity patch that moved
steadily East into a weak diffuse, barely visible, white-polarity
patch.
\revpar

FAF-1 exhibited the strongest and most \acp{FAF}-like flaring, at
08:05~UT.
It did so not only in \acp{AIA} 1600\,\AA\ but also in the \Halpha\
core, appearing as a very bright microflare.
At about that moment, filamentary arch-shaped brightenings extended
from it and connected to two others, FAF-2 to the North-East and
another \acp{FAF} to the South-West. 
Both then brightened in tandem.
EB-2 also flared again at 08:05~UT (Fig.~\ref{fig:lightcurves_EB2}),
but inspection of the \acp{AIA} 1600\,\AA\ movie does not suggest a
filamentary connection from FAF-1.

The double filamentary arches extended at apparent speeds over the
surface of 200\,\kms\ and then around 08:10~UT gave the impression of
lifting off upward in the form of a bright thin arc-shaped thread, or
giving such an appearance from successive brightening of a fan of
adjacent higher-up field loops, over a length of about 20~arcsec and
with an apparent projected rise speed of about 40\,\kms.
This apparent lift-off was vague in \acp{AIA} 1600\,\AA\ but produced a very
bright arch besides FAF-1 in \acp{AIA} 304, 171, and 193\,\AA\ during four
minutes, with the same morphology in these diagnostics.  
Thus, in contrast to \acp{EB}s which do not excite response in the hotter
\acp{AIA} channels, FAF-1 did so very markedly.

Inspection of the \acp{IRIS} spectra showed no spectacular profiles
from this million-Kelvin arch: the \SiIV\ lines were enhanced but
single-peaked, the \CII\ lines enhanced and double-peaked but fairly
average in width, \MgIIhk\ were enhanced and broadened, and there was
no sign of the \MgII\ triplet lines.
However, the \OIV\ lines at 1399 and 1401\,\AA\ were clearly present
throughout the hot arch.
These signatures merit further investigation, but studying and
displaying such hot coronal \acp{FAF} aftermaths falls outside the
scope of this \acp{EB} paper.
We concentrate here on the lower-atmosphere signatures at the FAF-1
site.

Unfortunately, the \acp{IRIS} slit did not sample FAF-1 at the time of
its \Halpha\ flaring and filamentary extension.
We therefore display the imaging at this time in the fourth row of
Fig.~\ref{fig:cutouts_FAF1} and show spectra from all slit passes
before and after.
In the fourth row the \Halpha\ core shows its microflare while the
onset of the filamentary extensions is clearest in the 1330\,\AA\
slitjaw image, but also noticeable in \Halpha.
There was no enhancement of the \Halpha\ wings, nothing like an
\acp{EB}, at that time.

However, the first row of Fig.~\ref{fig:cutouts_FAF1} displays
brightness patches like an \acp{EB}, but with the abnormality of also
appearing in the \Halpha\ core.
The latter is not very bright at this initial sampling time in in
Fig.~\ref{fig:lightcurves_FAF1}, but this is due to the presence of an
extended dark fibril within the wide integration contour.
The corresponding blue profiles in Fig.~\ref{fig:spectra_FAF1} appear
\acp{EB}-like, rather like EB-1's violet profiles in
Fig.~\ref{fig:spectra_EB1} but higher, also in \MgIIhk.
The blue \MgII\ triplet profile shows very high peaks, also as for the
violet sampling of EB-1.

The next sampling (second row, violet pixel and violet profiles)
produced bright \SiIV\ and \CII\ lines.
They remain less intense than the brightest from EB-2, but show
extremely wide and much blueshifted profiles, top raggedness, and deep
\NiII\ blends.
The deep self-absorption dips of the \CII\ lines remain at their rest
wavelengths but the rest of the profiles are so much blueshifted, as
are the \SiIV\ lines, that almost no red peaks remain.
The intensity ratio of the \SiIV\ lines still suggests thin \rev{to
thinnish} formation.
\rev{Both \SiIV\ lines resemble the \SiIV\ profiles for B-2, 3 and 4
in \citetads{2014Sci...346C.315P}, 
while the \CII\ lines are most alike the \CII\ lines for B-3 and 4,
suggesting that those three \acp{IB}s may in fact have been FAFs.}

The striking shape similarities between the \CII\ and \SiIV\ profiles
blueward from the nominal line centers and in the red tails suggest
similar sampling of FAF-1 in these parts of the \CII\ lines, \ie\ that
the profile structure is mostly set by similar effects of
Dopplershift, \rev{respectively} on the emissivity in optically-thin
formation and on \revpar the optical depth scales in optically-thick
formation.

The non-shifted \CII\ core dips suggest formation (likely scattering)
in a non-disturbed overlying region.
The violet \MgIIhk\ profiles also show wide wing extensions, with the
near equality of the two lines again suggesting optically thick
formation.
The violet \MgII\ triplet profile is about normal (nearly absent).
The \MnI\ blends in the blue \hk\ wings are deep dips.

The next sampling, again 8.6~min later but still 3~min before the
\Halpha\ microflare, produced the extraordinary orange profiles in
Fig.~\ref{fig:spectra_FAF1}.
The \SiIV\ and \CII\ lines have lower intensities, but are much wider
and more symmetrical, as if smeared by enormous thermal broadening or
sampling a wide distribution of very fast motions. 
They are appreciably blueshifted, but have extended red tails.
The \CII\ lines again appear very similar to the \SiIV\ lines, except
for the little dips at their nominal line centers suggesting minimal
absorption (scattering) in undisturbed gas along the line sight.
There are no absorption blends whatsoever.
The profile tops are ragged.

The orange \MgIIhk\ profiles reach as high intensities as the
brightest from EB-2 and with similar profiles but show only very small
central dips, with ragged appearance.
They are the only \hk\ sampling in all our spectra with obviously
unequal wing intensities between the two lines.
The cores share in this behavior. 
The ratio is about 1.5. 
The orange \CII\ profiles show similar difference.
It suggests that these lines sampled FAF-1 less thickly than our other
\acp{EB} and \acp{FAF} measurements.

The \MgII\ triplet lines show slight emission. 
Together, the orange profiles suggest sampling a very hot rising event
with much internal motion that was similarly sampled by the \SiIV,
\CII, and \hk\ lines, with only a small amount of undisturbed gas
causing dips at the \CII\ and \hk\ line centers.
It indeed seems an event on its way to become a bright million-Kelvin
feature in the \acp{AIA} diagnostics, becoming thin \rev{or thinnish} even
in \MgIIhk.

The next \acp{IRIS} sampling came 6~min after the \Halpha\ microflare (fifth
row, red pixel, red profiles) and while the bright million-Kelvin \acp{AIA}
arch was present to the North-East (directly above FAF-1 it was at a
height of about 6000~km and appeared projected well beyond the line of
sight to FAF-1).
These profiles are rather like the orange ones, but have lower
intensity and are less blueshifted.
They still have red tails but longer blue ones, show small \CII\
line-center dips, have deep \NiII\ blends but no \FeII\ blends, only
very weak \MnI\ blends, and ratios departing from optical thickness.
There was no specific brightening in the image cutouts anymore, except
the \Halpha\ core which shows a long post-flare tail in
Fig.~\ref{fig:lightcurves_FAF1}.
In summary, the red profiles suggest a cool-down aftermath.

In the last row of Fig.~\ref{fig:cutouts_FAF1} we also show the scene
in the next \acp{IRIS} sampling, in order to illustrate that the show
was over.
There was nothing of interest in the spectra (not shown) anymore.

\subsection{Details for FAF-2}

FAF-2 occurred close to FAF-1 but brightened earlier.
It also showed double filamentary extensions in 1600\,\AA\ and also
seemed to excite response in the hotter \acp{AIA} channels: slender
bright arches during 08:01:30--08:03:30~UT that were clearest in 171
and 193\,\AA\ and showed up as bright narrow arcs hovering over FAF-1
in projection to the North-East, but much less prominent than the
bright arch after FAF-1.

Figures~\ref{fig:cutouts_FAF2}--\ref{fig:spectra_FAF2} display our
standard-format results for FAF-2 in four successive \acp{IRIS}
samplings.
The first sampling (top row of Fig.~\ref{fig:cutouts_FAF2}, first
vertical time marker in Fig.~\ref{fig:lightcurves_FAF2}, no spectra
since not of interest) shows the scene seven minutes before FAF-2
brightened: nothing special.
The second and third samplings bracket its largest brightness; in the
fourth it had diminished (Fig.~\ref{fig:lightcurves_FAF2}). 
The \Halpha\ core showed fibrilar brightening parallel
to the filamentary extensions at 1600\,\AA\ but not co-spatially.

The blue profiles in Fig.~\ref{fig:spectra_FAF2} \rev{ are the widest
of all our specimens and} have deep absorption blends (also from \SI\ 
and \MnII).
They even show central dips in the \SiIV\ lines.
In this event the \MgIIhk\ lines share fully in the \SiIV\ and \CII\
behavior.
The similarity between the blue profiles of these six lines is
striking, except that the peak ratios reverse from blue-over-red for
\MgIIhk\ and the \CII\ lines to red-over-blue for the \SiIV\ lines.

The central dips of the \SiIV\ lines are not due to optically thick
self-absorption because they do not have the same absolute intensity
but still share the probability-corrected value.
They suggest absorption (scattering) in stationary, likely
undisturbed, gas along the line of sight that reaches sufficient
opacity for thick self-absorption scattering in the \CII\ and \MgIIhk\
lines.
Outside these cores the six lines exhibit strikingly similar profile
shapes.

The blue profiles are not only the brightest but also display the
largest raggedness. 
Inspection of the original spectral images suggests that most of the
deeper narrow dips are unidentified blends (for example the dips just
red and blue of the \NiII\ line between the \CII\ lines).
The \MnI\ and \MnII\ lines in the \hk\ wings are also strongly
present, in this case with significant blueshifts. 

The later samplings (violet and red) show roughly similar profiles at
diminishing intensities.
The violet sampling shows peak redshift for the \SiIV\ lines and
corresponding red-over-blue peak asymmetry for the \CII\ lines, but
not for \MgIIhk. 
The yet later red sampling shows small peak blueshift in the \SiIV\
lines, but similar slight blue-over-red \CII\ and \MgIIhk\ peak
asymmetries and similar line-center redshifts in these doublets (also
in the \Halpha\ core).
The blend blueshifts (also of the \MnI\ lines) became smaller than in
the blue sampling.
The \MgII\ triplet lines show up in absorption in all three samplings.
All profiles show some core raggedness.

In summary, FAF-2 showed yet wider profiles than FAF-1 but with more
symmetry and with also the \MgIIhk\ wings taking part.  
The later samplings again suggest a cool-down phase, but with more
undisturbed gas along the line of sight because there was no
blend-free sampling as the orange profiles of FAF-1 in
Fig.~\ref{fig:spectra_FAF1}; in contrast, blends including the \MnI\
lines were present in all three samplings, strongly so in the first
(blue) and last (red) ones.
We speculate that our slanted viewing passed through undisturbed gas
at larger height than the \acp{FAF} samplings because fibrilar
canopies tend to rise steeply away from network (containing the
reconnecting \acp{MC}s) to quieter internetwork areas.

\section{Discussion}\label{sec:discussion}

\subsection{\acp{EB} and \acp{FAF} properties}
\acp{EB}s and \acp{FAF}s show interesting similarities and
differences.
Both phenomena occur in emerging active regions and both probably mark
reconnection.
\acp{EB}s do this for reconnection of strong near-vertical fluxtubes
in the photosphere and appear as upright flames that remain under the
fibrilar chromospheric canopy \rev{(cf.~\PaperI, \PaperII, further
references in \citeads{2013JPhCS.440a2007R})}, 
or even below the upper photosphere when weak (EB-a, b, c). 
\acp{FAF}s show distinctive fibrilar morphology and are likely
reconnection events along the curved fields that define the canopy.

FAF-1 may have started below the canopy (but was then bright in the
\Halpha\ core, therefore formally not an \acp{EB}) but broke through
and even became an \Halpha-core microflare at 08:05~UT, two minutes
after it showed its widest ultraviolet profiles (orange in
Fig.~\ref{fig:spectra_FAF1}).   

FAF-2 seemed to reach less high since the blends and line-center dips
in all samplings in Fig.~\ref{fig:spectra_FAF2} suggest undisturbed
cooler stationary gas along the line of sight.

Both \acp{EB}s and \acp{FAF}s show outspoken bi-directional jet
signatures.
For the \acp{EB}s this is obvious since the \acp{IRIS} spectra of
EB-1, the onset of EB-2, and of EB-a, b, and c all show them directly,
spatially separated for the first two but mixed up along the line of
sight for the other three.
The bi-directional nature of the \acp{FAF}s is more complex because
they are less aligned with the line of sight (since also seen in the
images as rapid filamentary extensions) and cover larger temperature
ranges.
The striking difference between the violet sampling of FAF-1 and the
blue sampling of FAF-2 in the \acp{IRIS} profiles may simply be that
the first sampled only one of such jets, the other both. 
The \acp{IRIS} profiles from B-1 of
\citetads{2014Sci...346C.315P} 
then fit in this picture by showing larger Doppler separation.

\subsection{\acp{EB} temperature estimation}
The first conclusion from observing \acp{EB}s in the \acp{IRIS} lines
is that \acp{EB}s get very hot, especially in their tops.
Such apparent heating to very high temperature in \acp{IB}s was the
main message of \citetads{2014Sci...346C.315P} 
who wondered whether \acp{IB}s are \acp{EB}s or not.
Our results show that both \acp{EB}s and \acp{FAF}s can produce
\acp{IB} signatures that suggest exceedingly hot events.

How hot precisely?
\citetads{2014Sci...346C.315P} 
cited the coronal-equilibrium presence temperature of 80,000~K for
\SiIV, but remarked that the actual ion distribution may be closer to
\acp{LTE} \rev{and peak at lower temperature}.
\revpar

\begin{figure}
  \centerline{\includegraphics[width=0.8\columnwidth]{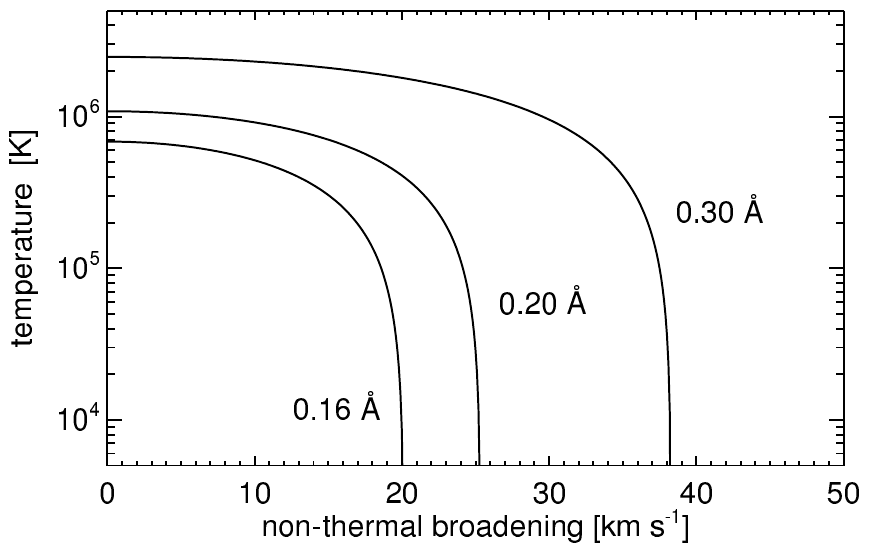}}
  \caption[]{\label{fig:fwhm} 
  Gaussian profile fitting for \SiIV~1403\,\AA.
  Each curve shows the combinations of temperature and non-thermal
  broadening that would produce the specified FWHM values
  \rev{bracketing our observations (0.16\,\AA\ for quiet Sun,
  the range 0.2--0.3\,\AA\ for EBs).}
  The instrumental broadening was set at 4.1\,\kms\ following
  \citetads{2014SoPh..289.2733D} 
  and \citetads{2014Sci...346A.315T}. 
  }
\end{figure}

\rev{Another way to estimate formation temperature is profile
matching.}
\revpar
In the idealized case of a convolution of thermal broadening with
Gaussian non-thermal broadening and instrumental broadening the
full-width at half maximum (FWHM) of the emergent profile from an
optically thin homogeneous feature is
$ \mbox{FWHM} = 1.67\,(\lambda/c) \, \sqrt{2kT/m + \xi^2 + \sigma^2}$
with $m$ the atomic mass, $\xi$ the non-thermal broadening, and
$\sigma$ the instrumental broadening.
\revpar
\rev{Figure~\ref{fig:fwhm} shows the possible combinations of thermal
and non-thermal broadening that reproduce our observed halfwidth values.
These range so wide that without precise knowledge of the non-thermal
contribution this approach fails.  In addition, our \acp{EB}s are not
truly thin nor homogeneous.}

\revpar

A \rev{better} way to establish \acp{EB} temperatures is profile matching
with more sophisticated modeling.
Modeling of \Halpha\ profiles from \acp{EB}s has been attempted by
\citetads{1983SoPh...87..135K}, 
\citetads{2010MmSAI..81..646B}, 
\citetads{2013A&A...557A.102B}, 
and
\citetads{2014A&A...567A.110B}. 
They all applied ad-hoc perturbations of a static standard model to
reproduce observed \Halpha\ moustaches.
Most defined the perturbation to not extend high, in order to avoid
non-observed brightening of the \Halpha\ core, but
\citetads{2014ApJ...792...13H} 
recognized that the core is actually formed in an overlying fibrilar
canopy and should not be modeled as an \acp{EB} property.
They applied a two-cloud fitting model, one for the \acp{EB}, the
other for the canopy.
These fitting exercises all claim that \acp{EB}s represent temperature
enhancements of the low standard-model chromosphere by at most a few
thousand Kelvin, usually less.
It seems highly unlikely that \SiIV\ lines as displayed here can be
obtained from any of them.

In addition to these trial-and-error fits, numerical \acp{EB}
simulations have been reported by
\citetads{2001ChJAA...1..176C}, 
\citetads{2007ApJ...657L..53I} 
expanding on the serpentine emergence simulation of
\citetads{1992ApJS...78..267N}, 
\citetads{2009A&A...508.1469A}, 
and most recently by \citetads{2013ApJ...779..125N}. 

The most extensive \revpar simulation is the one by
\citetads{2009A&A...508.1469A} 
who specifically targeted \acp{EB}s by setting up strong-field U-loop
emergence.
\revpar
It also delivered temperature enhancements of order 1000\,K.
However, there was no proper accounting for radiation and no spectral
synthesis of \eg\ \Halpha.

The most recent simulation, by
\citetads{2013ApJ...779..125N}, 
did not set up an active-region emergence event but quiet-Sun
magnetoconvection.
If their reconnection event was an \acp{EB} then \acp{EB}s should
appear all over the Sun, contrary to observation.
The continuum brightening, core brightening (``line gap'') of \FeI\
6303\,\AA\ and \Halpha\ inner-wing brightening in their synthesized
spectra (at top-down viewing, not slanted) suggests that they
simulated a pseudo-EB. 
In any case, the heating was less than 1000\,K.

The conclusion must therefore be that the \revpar fact that \acp{EB}s show
up in the \acp{IRIS} lines contradicts all \acp{EB} modeling efforts
so far.  

\begin{figure}
\centerline{\includegraphics[width=\columnwidth]{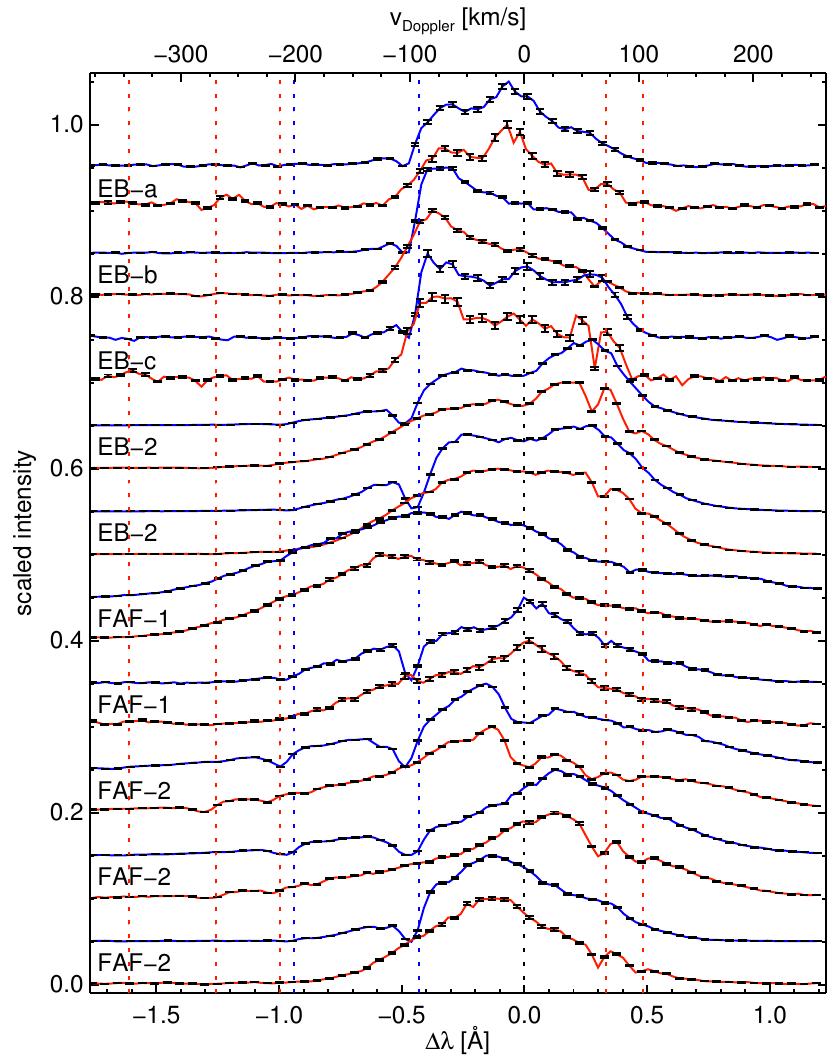}}
\caption[]{\label{fig:stagger} 
Stagger plot of normalized \SiIV\ profiles to inspect commonality.
For each feature that shows profile raggedness (labeled along the left
side) selected \SiIV\ 1394\,\AA\ profiles are shown in blue, the
corresponding \SiIV\ 1403\,\AA\ profiles in red, on a common
wavelength offset scale with corresponding Dopplershifts along the
top.
\rev{Error bars for Poisson-noise estimates are overplotted in black for
every other data point.}
The intensity range per line is 0.0--0.1, with offsets 0.05.
Common dips in same-color profiles are due to absorbing blend lines.
Common dips or humps per blue-red feature pair away from
$\Delta \lambda \is 0.0$ indicate fine-structure mapping.
Other fine structure is mostly measurement noise.
The dashed lines, with corresponding color coding, specify rest
wavelengths of major lines, from left to right: \OIV\ 1401.16\,\AA\
(red, \ie\ blend on \SiIV\ 1403\,\AA), \SI\ 1401.51\,\AA\ (red),
\FeII\ 1401.77\,\AA\ (red), \FeII\ 1392.82\,\AA\ (blue, \ie\ blend on
\SiIV\ 1394\,\AA), \NiII\ 1393.32\,\AA\ (blue), \SiIV\ 1394 and
1403\,\AA\ (black, defining $\Delta \lambda \is 0$), \FeII\
1403.10\,\AA\ (red), \FeII\ 1403.26\,\AA\ (red).
}
\end{figure}

\subsection{Fine structure in \acp{IRIS} profiles}
There is fine-structure raggedness in the green and orange EB-2
profiles in Fig.~\ref{fig:spectra_EB2}, the EB-a, b and c profiles in
Fig.~\ref{fig:spectra_EBabc}, the orange and red profiles of FAF-1 in
Fig.~\ref{fig:spectra_FAF1}, and the FAF-2 profiles in
Fig.~\ref{fig:spectra_FAF2}.

It can simply be measurement noise, especially at low count values
(\eg\ in the continua and overlapping wing part of \MgIIhk\ in
Fig.~\ref{fig:spectra_EB1}), but such noise contribution is smaller in
bright line cores.
The largest counts in the \SiIV\ lines were reached for EB-2
(Fig.~\ref{fig:spectra_EB2}), but also these cores show ragged fine
structure.

If the raggedness is not noise then it is either due to blends, which
when present should re-occur at the same wavelength in different
spatio-temporal samplings of a given line, or to profile mapping of
fine structure in the event along the line of sight.
In the latter case and in optically thick line formation, similar
small-scale profile structure should appear at comparable optical
depth sampling in different lines per pixel sampling, \ie\ slightly
further out from line center in the stronger component of a doublet.
In optically thin line formation similar profile structure should
occur at about the same Dopplershift in different lines per sampling.

Figure~\ref{fig:stagger} details this choice for the two \SiIV\ lines. 
In all cases their height ratio suggests optically thin \rev{or
thinnish} line formation, so that the choice to explain raggedness is
between noise, blends, and Doppler mapping of fine structure.

Most larger narrow dips re-occur at the same wavelength in different
samplings of the same line (either blue or red) and are therefore
blends. 
The major ones are identified by dashed lines with corresponding line
colors. 
Most of these are blueshifted over about 10\,\kms.
There are more blends, especially in the FAF-2 profiles as noted
above.
\revpar

The EB-a and EB-c profiles in Fig.~\ref{fig:stagger} show the largest
core raggedness, but also the largest raggedness in the adjacent
continua \rev{and the largest noise (error bars in 
Fig.~\ref{fig:stagger})} \revpar
because these cores reach only
about 200 and 100 counts, respectively (Fig.~\ref{fig:spectra_EBabc}).
Indeed, the raggedness appears larger for the weaker red profiles and
smaller in the brighter profiles from EB-b.

Finally, the somewhat larger-scale deviations from Gaussian profiles
in the \SiIV\ cores in Fig.~\ref{fig:stagger} often show good
similarity between the red and blue pair of profiles per sample.
In view of the apparent optically-thin \rev{(or thinnish)} line
formation of the \SiIV\ lines we attribute such common fine structure
to Dopplershift mapping of features along the line of sight.

For the \acp{EB}s the presence of Dopplershift fine-structure fits the
appearance of \acp{EB}s in our \acp{SST} \Halpha\ movies as rapidly
flickering flames and their bi-directional jet signatures. 
We surmise that rapid successive, intermittent reconnection of newly
arriving opposite-polarity concentrations with varying flux content
causes fast variations that eventually end up as fine structure in the
resulting hot events.

For the \acp{FAF}s the wide extent of the profile tails agree with the
apparent observed filament extension speeds of 200~\kms.
The picture of thin fast-rising heating events also fits the
morphology of the apparent lift-off of thin arcs and resulting bright
arches in \acp{AIA} 304, 171 and 193\,\AA, in particular after FAF-1. 

\subsection{Nature of the hot events}
EB-2 and the two \acp{FAF}s produced spectra with \acp{IB} hot-event
signatures.
EB-1 may have done so too, but its aftermath was not sampled by the
\acp{IRIS} slit.
The weaker EB-a, b and c did not. 
The upshot is that strong \acp{EB}s can do so.
Both \acp{FAF}s certainly did.

The production of a hot event by upward progressing reconnection is
not surprising.
When in an \acp{EB} two photospheric fluxtubes cancel against each
other by reconnection, one may expect local temperature increase below
a factor two because such fluxtubes tend to obey magnetostatic
equipartition with their surroundings.
This is indeed seen in the numerical \acp{MHD} simulation of
\citetads{2009A&A...508.1469A}. 
However, if the reconnection site then proceeds upward,
the heating ratio increases because the magnetic energy diminishes
less than the gas energy due to larger scale height.  
Much larger temperature increase may there be expected.
In addition, the jet kicks are likely to cause Alfv\'enic wave
generation and dissipation. 
Therefore, the observational indication that \acp{FAF} reconnection
occurs higher up than \acp{EB} reconnection may explain why \acp{FAF}s
produce larger \acp{IB}s and \rev{also} million-Kelvin arches as FAF-1 did.

Various scenarios come to mind to produce hot \revpar
gas with spectral blend superimposing by cooler and relatively 
stationary gas.
The first is that this gas is simply undisturbed upper-photosphere or
chromospheric canopy gas along the line of sight.
We suggest that this is the case for the \MnI\ blends in the \MgIIhk\
wings for the lines of sight to the feet of EB-1 and EB-2, and may
likewise signify small event heights for EB-a, EB-b, and EB-c. 
Such blends were not present in FAF-1 which had the largest
higher-atmosphere response of all our events, but for FAF-2 they were
prominently present in the blue sampling in
Fig.~\ref{fig:spectra_FAF2}, superimposed on the widest of all our
ultraviolet profiles, and still in the red sampling 17~min later.
Assuming that these were from undisturbed ``normal'' gas along the
line of sight implies that even FAF-2, which also showed filamentary
extensions and \acp{AIA} hot-diagnostic brightening, did not reach
very high in its \acp{IRIS} samplings.

\rev{We similarly speculate that the \FeII\ and \NiII\ blends on the \SiIV\
and \CII\ lines, which typically show blueshifts up to about 10\,\kms, as
also found by \citetads{2014Sci...346C.315P}, 
sample adjacent gas harboring upward propagating shocks as those
in internetwork regions
(\eg\ \citeads{1997ApJ...481..500C}; 
\citeads{2007A&A...473..625L}; 
\citeads{2009A&A...494..269V}) 
and in dynamic fibrils near network and plage
(\eg\ 
\citeads{2006ApJ...647L..73H}; 
\citeads{2007ApJ...655..624D}; 
\citeads{2007ApJ...666.1277H}; 
\citeads{2008ApJ...673.1194L}).} 

Alternatively, an EB-like start-up \acp{FAF} may kick cool
photospheric gas up to large height where it can cause line-center
dips and absorption blends, or these may originate in a post-bomb
cooling cloud. 
However, such \rev{scenarios} seem unlikely in view of the lack of large
Dopplershifts in the blend lines and central dips.

\subsection{\acp{EB} and \acp{FAF} visibility in \MgII\ 2798\,\AA}
In our \acp{IRIS} spectra the \MgII\ triplet lines appears with
interesting behavior.
Their emission profiles closely mimic the \CII\ line shapes in
\acp{EB}s and in the initial \acp{EB}-like stage of FAF-1, but not in
the \acp{FAF} spectra with stronger \acp{IB} signatures.
Strong presence of this line suggests a steep, deeply located
temperature rise
(\citeads{2015arXiv150401733P}). 
The marked appearances of \MgII\ 2798\,\AA\ in
Figs.~\ref{fig:spectra_EB1}, \ref{fig:spectra_EB2} and
\ref{fig:spectra_EBabc} are indeed in good agreement with the \CII\
and \SiIV\ indications of high temperature already in the lower-part
samplings.

In Fig.~\ref{fig:spectra_FAF1} the line is very strongly present in
the first sampling of FAF-1, almost as tall as \hk,
so that the startup of FAF-1 resembled a low-atmosphere \acp{EB} also
in this respect.
The later \MgII\ triplet profiles of FAF-1 in
Fig.~\ref{fig:spectra_FAF1} are less extraordinary.
In all FAF-2 samplings \rev{this} line appeared in absorption
(Fig.~\ref{fig:spectra_FAF2}).

\paragraph{\acp{EB} and \acp{FAF} visibility in \Halpha}
\Halpha\ is an extraordinary line, as is obvious from any solar
\Halpha\ filtergram. 
In \acp{EB}s and \acp{FAF}s it is also special.
In EB-1 the \Halpha\ top producing enhanced \CII\ and \SiIV\ emission
is not seen in the first panel of Fig.~\ref{fig:cutouts_EB1}, nor in
the second panel.
This suggests that hydrogen was already ionized in the upper part, as
one would expect from \SiIV\ visibility.
However, in the other rows of Fig.~\ref{fig:cutouts_EB1} and also in
Fig.~\ref{fig:cutouts_EB2} the \Halpha-wing morphology resembles the
\acp{IRIS} 1330\,\AA\ slitjaw images.
We believe that such hot response of \Halpha\ stems from severe
non-equilibrium recombination of hydrogen
(\citeads{2002ApJ...572..626C}; 
\citeads{2007A&A...473..625L}).

\section{Conclusions}  \label{sec:conclusions}

We have combined a comprehensive suite of solar observations.
The relatively large field of view, unsurpassed image quality, and
fast cadence of the \Halpha\ imaging spectroscopy with the \acp{SST}
was indispensable to recognize both \acp{EB}s and overlying fibrils
from their spatial, temporal, and spectral behavior. 
The full-time full-disk monitoring by \acp{SDO} served to separate
\acp{EB}s and \acp{FAF}s in \acp{AIA} 1700 and 1600\,\AA\ images and
to inspect magnetic field evolution in \acp{HMI} magnetograms.
The \acp{IRIS} slitjaw images, effectively providing high-cadence
large-field images in the major \acp{IRIS} lines, are an extremely
valuable asset that previous solar ultraviolet spectrometers did not
furnish.
And last but not least, the spectra in the well-chosen set of
\acp{IRIS} lines emerge as varied ``unveiled'' \acp{EB} diagnostics
not hampered by overlying fibrils and offering rich signatures of what
happens in the solar atmosphere at \acp{EB} and \acp{FAF} sites.
They testify to Pannekoek's dictum that ``spectra constitute the
astronomer's treasure chest''.

We summarize our conclusions for \acp{EB}s as follows:

\leftmargini=3ex
\begin{enumerate}  
\itemsep=-0.2ex \vspace{-1.0ex}

\item the cores of \Halpha\ and the \MgII\ \hk\ lines sample
overlying chromospheric fibrils that are unaffected by the underlying
\acp{EB}.
In these lines \acp{EB}s are visible only well away from line center;

\item the \acp{IRIS} \MgII\ triplet, \CII, and \SiIV\ lines sample the
Ellerman bomb itself, often with optically-thin \rev{or near-thin}
formation of the \SiIV\ cores;

\item these ``unveiled'' \acp{IRIS} diagnostics indicate that the tops of
Ellerman bombs become much hotter than all previous estimates in the
literature; 

\item they also give direct evidence of bi-directional jet behavior,
with downdrafts of the lower parts and faster updrafts of the hotter
upper parts;

\item subsequently, very hot post-bomb gas appears with wide and
complex ultraviolet line profiles that suggest large Dopplershifts,
possibly still from bi-directional jets, and much fine-scale
structure. 
Even these remain a sub-canopy phenomenon. 

\end{enumerate}

\noindent
\acp{FAF}s seem to represent a comparable reconnection phenomenon, but
breaking through or progressing along the chromospheric canopy and
causing much hotter structures that also become evident in
million-Kelvin \acp{AIA} diagnostics.
The main difference with \acp{EB}s seems that the reconnection is
located or proceeds higher, but the blends and line-center-dips in the
\acp{IRIS} profiles from FAF-2 still suggest fairly deep formation.

For future \acp{EB} modeling the ultraviolet line profiles from
\acp{IRIS} represent highly varied diagnostics furnishing such rich
detail that modeling which succeeds in good reproduction is bound to
be close to correct.
If it does not succeed then such failure is bound to be instructive
also.
In this manner, \acp{EB}s are likely to become the first solar
reconnection phenomenon for which detailed modeling may be verified
with certainty.

\revpar

Since \acp{FAF}s seem of larger interest with regards to
upper-atmosphere mass and heat loading, verified \acp{EB} modeling
seems a worthwhile stepping stone to modeling \acp{FAF}s properly.
A good example in this direction is the recent study by
\citetads{2014ApJ...788L...2A} 
of the formation of small flares from strong-field magnetoconvection
producing serpentine emergence of the type proposed for \acp{EB}s.  
Their resulting heating events and coronal jets are more substantial
and located higher than \acp{EB}s, but may well describe what we have
called \acp{FAF}s here.

Observationally, the next step is easier: catch \acp{EB}s and
\acp{FAF}s in joint \acp{SST} and \acp{IRIS} (and of course \acp{SDO})
observing campaigns targeting emerging active regions well away from
disk center with a faster \acp{IRIS} repeat cadence than in datasets 2
and 3.
It would also be good to roll \acp{IRIS} to put its slit along the
projected vertical per target.

\acknowledgments 

Our research has been partially funded by the Norwegian Research
Council and by the \acl{ERC} under the European Union's Seventh
Framework Programme (FP7/2007-2013)\,/\,ERC grant agreement
nr.~291058.
B.D.P. was supported by NASA contract NNG09FA40C (IRIS).
\rev{B.D.P., M.C., and L.R.v.d.V. have}
benefited from discussions at the \acdef{ISSI} meeting on ``Heating of
the magnetized chromosphere''. 
IRIS is a NASA small explorer mission developed and operated by LMSAL
with mission operations executed at NASA Ames Research Center
and major contributions to downlink communications funded by the
Norwegian Space Center through an ESA PRODEX contract.
The SST is operated on the island of La Palma by the Institute for
Solar Physics of Stockholm University in the Spanish Observatorio del
Roque de los Muchachos of the Instituto de Astrof{\'\i}sica de
Canarias.
We thank T.~Golding, T.~Pereira, and H.~Skogsrud for assistance with
the SST observations and the referee for many useful suggestions.

\mbox{} 


\end{document}